\documentclass[journal,draftclsnofoot,onecolumn,12pt]{IEEEtran}

\usepackage{xspace,amsmath,amssymb,amsfonts,epsfig,subfigure,syntonly}
\usepackage{cite,bm,color,url,textcomp,empheq,boxedminipage}
\usepackage{algorithmicx,algorithm}
\usepackage{epstopdf,makecell}
\usepackage{empheq}
\usepackage{pifont}

\usepackage{graphicx,graphics}  
\usepackage{multirow,multicol}
\usepackage{psfrag}    
\usepackage{stfloats}
\usepackage{url}

\newtheorem{lemma}{\underline{Lemma}}[section]

\newtheorem{proposition}{\underline{Proposition}}[section]

\newtheorem{remark}{\underline{Remark}}[section]

\newcommand{\mv}[1]{\mbox{\boldmath{$ #1 $}}}

\long\def\symbolfootnote[#1]#2{\begingroup
\def\thefootnote{\fnsymbol{footnote}}
\footnote[#1]{#2}\endgroup}

\psfull

\allowdisplaybreaks[4]

\begin{document}
\title{Common Throughput Maximization for UAV-Enabled Interference Channel with Wireless Powered Communications}
\author{Lifeng~Xie,~\IEEEmembership{Student Member,~IEEE,} Jie~Xu,~\IEEEmembership{Member,~IEEE,} and Yong~Zeng,~\IEEEmembership{Member,~IEEE}
\thanks{Part of this paper has been presented at the IEEE International Conference on Communication Systems (ICCS) 2018, Chengdu, China, December 19--21, 2018\cite{conferenceversion}. {\it(Corresponding author: Jie Xu.)}}
\thanks{L. Xie and J. Xu are with the School of Information Engineering, Guangdong University of Technology, Guangzhou, China. (e-mail: lifengxie@mail2.gdut.edu.cn, jiexu@gdut.edu.cn).}
\thanks{Y. Zeng is with the National Mobile Communications Research Laboratory, School of Information Science and Engineering, Southeast University, Nanjing, China, and also with the Purple Mountain Laboratory, Nanjing 211111, China. (e-mail: yong\_zeng@seu.edu.cn).}
}
\maketitle

\vspace{-20pt}\begin{abstract}
This paper studies an unmanned aerial vehicle (UAV)-enabled two-user interference channel for wireless powered communication networks (WPCNs), in which two UAVs wirelessly charge two low-power Internet-of-things (IoT)-devices on the ground and collect information from them. We consider two scenarios when both UAVs cooperate in energy transmission and/or information reception via interference coordination and coordinated multi-point (CoMP), respectively. For both scenarios, the UAVs' controllable mobility is exploited via trajectory design to not only enhance the wireless power transfer (WPT) efficiency in the downlink, but also mitigate the co-channel interference for wireless information transfer (WIT) in the uplink. In particular, the objective is to maximize the uplink common (minimum) throughput of the two IoT-devices over a finite UAV mission period, by jointly optimizing the trajectories of both UAVs and the downlink/uplink wireless resource allocation, subject to the maximum flying speed and collision avoidance constraints at UAVs, as well as the individual energy neutrality constraints at IoT-devices. Under both scenarios of interference coordination and CoMP, we first obtain the optimal solutions to the two common-rate maximization problems in well structures for the special case with sufficiently long UAV mission duration. Next, we obtain high-quality solutions for the practical case with finite UAV mission duration by using the alternating optimization and successive convex approximation (SCA).
\end{abstract}

\vspace{-15pt}\section{Introduction}
Wireless powered communication network (WPCN) has emerged as a promising technique to provide ubiquitous wireless energy and data access for a massive number of low-power devices in the upcoming Internet-of-things (IoT) era by unifying both wireless power transfer (WPT) and wireless information transfer (WIT). In WPCN, dedicated access points (APs) are deployed to broadcast radio-frequency (RF) signals to wirelessly charge IoT-devices in the downlink, which then utilize the harvested wireless energy to send information to the APs in the uplink. It is envisioned that the WPCN can achieve in principle perpetual operation for future IoT networks\cite{WPTbooks,WPTBCRZ,YongWPT,Bi2015,WangFeng,WPCNsurvey,Rui2013WPCN,WPCNrelay,WPCNpower,SUMRATEWPCN,MIMOWPCN}. Recently, extensive research on WPCNs has been conducted under different setups such as multiuser systems \cite{Rui2013WPCN}, relaying systems \cite{WPCNrelay}, and interference channels \cite{SUMRATEWPCN}, in which techniques such as multi-antenna beamforming \cite{MIMOWPCN}, power control \cite{WPCNpower}, and user cooperation \cite{Usercooperation} were used to enhance the performance of both downlink WPT and uplink WIT. However, these prior works mainly focused on conventional WPCNs with APs deployed at fixed locations on the ground, in which both the WPT and WIT performance are fundamentally limited by the severe signal propagation path loss, especially when the distances between APs and IoT-devices become large.

Recently, motivated by the success of UAVs in abundant applications (e.g., cargo delivery, aerial surveillance, filming, and especially  wireless communications \cite{C.Zhan,RIB,WuQing,UAVSWIPT,walidsaad,ZYUAVsurvey,WSUAV,Q.WU,YUXU,BinLi1,BinLi2} and WPT \cite{JieXuWPT,Jiewpt,Yundi,yulin}), UAV-enabled WPCN has attracted growing research interests. Specifically, UAVs can be dispatched as a new type of aerial APs for wirelessly charging low-power IoT-devices on the ground, and collecting information from them at the same time. Compared to conventional WPCN with fixed APs on the ground, UAV-enabled WPCN can be more swiftly deployed to cover larger areas under critical situations (e.g., after natural disasters). Besides, by exploiting the UAVs' fully controllable mobility, UAV-enabled aerial APs can properly adjust their locations over time (a.k.a. trajectories) to shorten the distances to their serving IoT-devices, thus enhancing the wireless coverage and improving both WIT and WPT efficiency\cite{Korea2,Xie,WPCNnew}.

In the literature,
the authors in  \cite{Xie} studied the UAV-enabled multiuser WPCN with one UAV-enabled AP serving multiple IoT-devices on the ground. By considering the time-division-multiple-access (TDMA) protocol for WPT and WIT, the UAV optimizes its 2D trajectory together with the time and power allocation over time, to maximize the common (or minimum) throughput of those IoT-devices, subject to their individual energy harvesting/neutrality constraints. In particular, the authors in \cite{Xie} proposed a successive hover-and-fly (SHF) trajectory design, in which the UAV hovers at a set of optimized locations over time for WPT and WIT with different users, respectively, and successively visiting these locations at the maximum speed. Furthermore, \cite{WPCNnew} considered the positioning optimization for the UAV-enabled WPCN, in which the UAV is positioned to hover at one single optimized location over the whole transmission duration for efficient WPT and WIT. In addition, \cite{Korea2} investigated the trajectory design for the UAV-enabled WPCN with two UAVs serving multiple IoT-devices, where the two UAVs are used for only WPT and WIT, respectively (instead of for both at the same time).

Despite such a rapid research progress, the aforementioned prior works on UAV-enabled WPCN mainly focused on the scenario with one single UAV\cite{Xie, WPCNnew} or two UAVs for independent WPT and WIT\cite{Korea2}. In general, to further enhance the performance of WPCN (in terms of both coverage and transmission efficiency), it is important to dispatch multiple UAVs to cooperatively serve multiple IoT-devices with joint WPT and WIT design. In this case, the uplink WIT among multiple IoT-device-UAV pairs corresponds to a multiuser interference channel, in which the simultaneous transmission by different IoT-devices results in co-channel interference at the UAVs; while the downlink WPT from UAVs to IoT-devices corresponds to a distributed WPT system with multiple mobile energy transmitters (ETs), in which these ETs (UAVs) may collaborate with each other (via, e.g., collaborative energy beamforming\cite{Xu2014A,SLee}) for enhancing the WPT efficiency. This thus introduces new technical challenges in the multiple UAVs' trajectory optimization, which critically depends on the transmission/reception strategies at the UAVs and needs to follow distinct design principles for WIT and WPT, respectively. Take the interference coordination scenario with two UAVs transmitting/receiving independent (energy/information) signals as an example. For uplink WIT, multiple UAVs should fly far away from their non-serving IoT-devices to avoid interference \cite{WuQing}; for downlink WPT, by contrast, these UAVs may need to hover/fly above the center points among IoT-devices for simultaneously charging them\cite{JieXuWPT}. As such, how to optimize multiple UAVs' trajectories to maximize the performance of UAV-enabled interference channel with both downlink WPT and uplink WIT is an important but challenging task, especially due to the involvement of the individual energy neutrality constraints at IoT-devices as well as the flying speed and collision avoidance constraints at UAVs. This task, to our best knowledge, has not been well addressed in the literature yet.

\begin{figure}
  \centering
  \includegraphics[width=8cm]{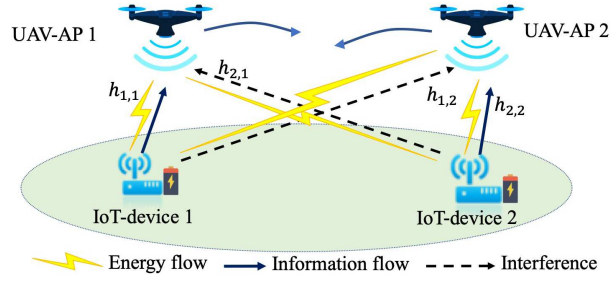}\\
  \caption{System model of the UAV-enabled two-user interference channel for wireless powered communications.}\label{fig1}
\end{figure}

To fill the above gap, in this paper we investigate a basic UAV-enabled two-user interference channel for WPCN as shown in Fig. \ref{fig1}, in which two UAVs are dispatched as flying APs to charge two ground IoT-devices via WPT in the downlink, and each IoT-device relies on its harvested RF energy to send information to its corresponding UAV in the uplink. In particular, we consider two scenarios with interference coordination and coordinated multi-point (CoMP) transmission/reception between the two UAVs, respectively. Notice that interference coordination and CoMP are two widely adopted base station (BS)/AP cooperation techniques in wireless networks\cite{comp1,DAS,liuliang}, which can efficiently handle the co-channel interference without and with the requirement of data sharing among BSs/APs, respectively. Under the two different scenarios, we study how to exploit the two UAVs' cooperative trajectories design to maximize the system performance. The main results of this paper are summarized as follows.

\begin{itemize}
\item Under both interference coordination and CoMP scenarios, we aim to maximize the uplink common (minimum) throughput of the two IoT-devices over a particular UAV mission period, by jointly optimizing the trajectories of both UAVs and the downlink/uplink wireless resource allocations, subject to the UAVs' maximum flying speed and collision avoidance constraints, as well as the IoT-devices' individual energy neutrality constraints. The two formulated uplink common throughput maximization problems are both non-convex and thus difficult to be solved optimally.
\item To solve the common rate maximization problem under each scenario, we first consider the special case with a sufficiently long UAV mission duration. In this case, we obtain the solution to the formulated uplink common throughput maximization problem, which has an interesting multi-location-hovering structure, such that each of the two UAVs hovers at different locations over time for WPT and WIT. We show that the hovering locations for the scenario with interference coordination are distinct from those for CoMP.
    Next, we consider the general case with a finite mission duration. In this case, we propose an alternating optimization based algorithm to obtain an efficient solution, by using the techniques of the convex optimization and successive convex approximation (SCA).


\item Finally, we present numerical results to validate the performance of our proposed designs. It is shown that our proposed designs with joint UAV trajectory and wireless resource optimization significantly outperform benchmark schemes without trajectory optimization, and the employment of CoMP achieves much higher throughput than interference coordination. Furthermore, when the UAV mission duration becomes sufficiently long, our proposed alternating-optimization-based designs are shown to approach the corresponding performance upper bounds by the multi-location-hovering solutions under both scenarios.
\end{itemize}

Notice that this paper differs significantly from its conference version \cite{conferenceversion}. While \cite{conferenceversion} only adopted the alternating-optimization-based algorithm for maximizing the common throughput in the interference coordination scenario, this paper further considers the CoMP scenario and derives the insightful multi-location-hovering solutions to the formulated problems under both scenarios when the UAV mission durations becomes sufficiently large.

The remainder of this paper is organized as follows. Section \ref{sec2} introduces the system model of the UAV-enabled two-user interference channel for WPCN and formulates the uplink common throughput maximization problems. Sections \ref{sec:P2} and  \ref{sec6} present efficient algorithms to solve the formulated problems for the scenarios with interference coordination and CoMP, respectively. Section \ref{sec9} provides numerical results to validate the effectiveness of our proposed designs. Finally, Section \ref{sec:con} concludes this paper.

{\it Notation:} Scalars are denoted by lower-case letters, vectors by bold-face lower-case letters, and matrices by bold-face upper-case letters. 
For a non-singular square matrix $\mv{M}$, $\mv{M}^{-1}$ denotes its inverse. $\mathbb{E}[\cdot]$ denotes the statistical expectation. $\mathbb{C}^{x\times y}$ denotes the space of $x\times y$ complex-valued matrices. $\| \cdot \|$ denotes the Euclidean norm of a vector.

\vspace{-15pt}\section{System Model and Problem Formulation}\label{sec2}

As shown in Fig. 1, we consider a UAV-enabled WPCN, where two UAVs are dispatched to periodically charge two ground IoT-devices via WPT in the downlink, and each IoT-device $k\in\{1,2\}$ utilizes its harvested energy to send information to the corresponding UAV in the uplink. Suppose that each IoT-device $k$ has a fixed location $(x_k,y_k,0)$ on the ground in a three dimensional (3D) Cartesian coordinate system, where $\bold{w}_k =(x_k,y_k)$ denotes the horizontal coordinate of IoT-device $k$. Let $D$ denote the distance between the two IoT-devices. Without loss of generality, we assume that $x_1=-D/2$, $x_2=D/2$, and $y_1=y_2=0$. The IoT-devices' locations are assumed to be {\it a-priori} known by the UAVs to facilitate their trajectories design and wireless resource allocations.

We focus on a particular UAV mission period $\mathcal T \triangleq (0,T]$ with finite duration $T$ in second (s), in which the UAVs fly horizontally at a fixed altitude $H>0$ in meter (m). The whole period $\mathcal T$ is discretized into a finite number of $N$ time slots, each with equal duration $\delta=T/N$. Note that for each time slot $n\in \mathcal N \triangleq \{1,...,N\}$, we have the corresponding period as $\mathcal T_n = ((n-1)\delta, n\delta]$. Also note that the duration $\delta$ is chosen to be sufficiently small, such that the UAVs' locations can be assumed to be approximately unchanged within each slot $n$. Let $\bold{q}_m(t)=(x_m(t),y_m(t)),t\in\mathcal T$ denote the location of UAV $m\in\{1,2\}$ projected on the horizontal plane. Hence, at any time slot $n\in\mathcal N$, we have $\bold{q}_m(t)\approx{\bold {q}}_m[n]=(x_m[n],y_m[n]),t\in\mathcal T_n$. Furthermore, let ${\bold{q}}_m^{\text{initial}}$ and ${\bold{q}}_m^{\text{final}}$ denote the initial and final locations of UAV $m$, respectively. Then we have
\begin{align}
{\bold{q}}_m[0]={\bold{q}}_m^{\text{initial}},
{\bold{q}}_m[N]={\bold{q}}_m^{\text{final}},\forall m\in\{1,2\}.\label{IFC}
\end{align}
We assume that the UAV mission duration $T$ satisfies $T\ge \|{\bold{q}}_m^{\text{initial}}-{\bold{q}}_m^{\text{final}}\|/ \tilde{V}_{\text{max}},\forall m\in\{1,2\}$, with $\tilde{V}_{\max}$ denoting the maximum flying speed of each UAV, such that there exists at least one feasible trajectory for each UAV to fly from its initial location to final location. Suppose that $d_{\text{min}}$ denotes the minimum inter-UAV distance to avoid collision. Accordingly, the UAVs' trajectories are subject to the maximum speed constraints and collision avoidance constraints given by
\begin{align}
&\|\bold{q}_m[n+1]-\bold{q}_m[n]\|\le {V}_{\max}, \forall n\in\mathcal N\setminus\{N\},m\in\{1,2\},\label{speedcon}\\
&\|\bold{q}_1[n]-\bold{q}_{2}[n]\|^2\ge d^2_{\text{min}},\forall n\in\mathcal N,\label{conconcon}
\end{align}
where $V_{\max} = \tilde V_{\max} \delta$ denotes the maximum UAV displacement at each time slot. Accordingly, the distance between IoT-device $k\in\{1,2\}$ and UAV $m\in\{1,2\}$ at time slot $n\in\mathcal N$ can be well-approximated by a fixed value determined by their approximated locations:
\begin{align}
\bar d_{k,m}(t)\approx d_{k,m}[n]=\sqrt{\lVert\bold{q}_m[n]-\bold{w}_k\rVert^2+H^2},\forall t\in\mathcal T_n.
\end{align}

We assume that the wireless channels between the two UAVs and the two IoT-devices are dominated by the line-of-sight (LoS) links due to the short distances (e.g., several or tens of meters) for WPT\cite{WPTBCRZ,WPTbooks,YongWPT,Bi2015}, similar to in prior works \cite{Jiewpt,JieXuWPT,yulin,Yundi,Xie,WPCNnew,Korea2}. Furthermore, we notice that for each wireless link, the path loss changes at a much slower time scale than the phase change\cite{liuliang}. 
Therefore, we consider the free-space path loss model with random channel phases, where the channel coefficient between UAV $m\in\{1,2\}$ and IoT-device $k\in\{1,2\}$ at time slot $n\in\mathcal N$ is given by
\begin{align}
h_{k,m}[n]=\sqrt{\frac{\beta_{\rm 0}}{\lVert\bold{q}_m[n]-\bold{w}_k\rVert^2+H^2}}e^{j \theta_{k,m}[n]},\label{channel}
\end{align}
where $\beta_{\rm 0}$ denotes the channel power gain at a reference distance of $d_{\rm 0}=1$ m, $j =\sqrt{-1}$, and $\theta_{k,m}[n]$ denotes the random phase of the wireless channel that is assumed as a uniformly distributed random variable in $[0, 2\pi)$, and independent over time and different UAV-device pairs.\footnote{The assumption of the uniformly distributed random phases is explained as follows. At any slot $n$, we have $\theta_{k,m}(t)=\frac{2\pi}{\lambda}\bar d_{k,m}(t),\forall t\in\mathcal T_n$, where $\lambda$ denotes the signel wavelength that generally takes a very small value (e.g., $1/3$~m at the frequency of 900 MHz). In this case, $\theta_{k,m}(t)$ is critically dependent on $\bar d_{k,m}(t)$, and a slight deviation of $\bar d_{k,m}(t)$ from $d_{k,m}[n]$ may significantly change the value of $\theta_{k,m}(t)$. Therefore, we cannot simply approximate $\theta_{k,m}(t)$ as the constant $\frac{2\pi}{\lambda}d_{k,m}[n]$; instead we model $\theta_{k,m}[n]$ as a uniformly distributed random variable in $[0,2\pi)$.} We also assume that the downlink WPT and uplink WIT use the same frequency band but over orthogonal time instants. Therefore, due to channel reciprocity, at any time slot $n\in\mathcal N$, the wireless channels for both uplink WIT and downlink WPT are identical. Furthermore, we consider two scenarios with interference coordination and CoMP transmission/reception at UAVs, which are detailed in the following.

\vspace{-15pt}\subsection{Scenario with Interference Coordination}\label{sec:inter}

With interference coordination, each of the two UAVs sends independent energy signals to the ground IoT-devices and individually decodes the information signals received from its associated IoT-device. At each time instant, the two UAVs operate in the same transmission mode for either WPT or WIT.\footnote{Notice that at each time instant, there exists another possible transmission mode with one UAV used for WPT and the other used for WIT at the same time, which, however, is not considered in this paper. This is due to the fact that in this mode, the resulting interference from the WPT-UAV (for energy transmission) to the WIT-UAV (for information reception) will be too strong, as the transmit power for WPT at UAV is significantly higher than the transmit power for WIT at low-power ground IoT-devices.} Suppose that at each time slot $n\in\mathcal N$, the WPT and WIT of both UAV-device pairs are operated over two orthogonal sub-slots with duration $\delta_E[n]\ge 0$ and $\delta_I[n]\ge 0$, respectively, with $\delta_E[n]+\delta_I[n]\le \delta,\forall n\in\mathcal N.$

First, we consider the downlink WPT at time slot $n\in\mathcal N$ with duration $\delta_E[n]$, during which each UAV $m\in\{1,2\}$ transmits independent energy signals $s_m[n]$ for charging the two IoT-devices simultaneously. Suppose that each UAV $m\in\{1,2\}$ adopts a constant transmit power $P$. Then the harvested energy at IoT-device $k$ is given by
\begin{align}
E_k(\delta_E[n],\{\bold{q}_{m}[n]\})&=\sum_{ m=1}^2 \eta P\delta_E[n]g_{k,m}[n]=\sum_{m=1}^2\frac{ \eta P\beta_{\rm 0}\delta_E[n] }{\lVert\bold{q}_{m}[n]-\bold{w}_k\rVert^2+H^2},\label{eqn:linear:model}
\end{align}
where $0<\eta\le 1$ denotes the RF-to-direct current (DC) energy conversion efficiency at the energy harvester of each IoT-device and $g_{k,m}[n]\triangleq|h_{k,m}[n]|^2=\frac{\beta_{\rm 0}}{\lVert\bold{q}_m[n]-\bold{w}_k\rVert^2+H^2}$ denotes the channel power gain from UAV $m\in\{1,2\}$ to IoT-device $k\in\{1,2\}$ at slot $n\in\mathcal N$. Hence, the total energy harvested by IoT-device $k$ over the entire period with duration $T$ is given by
\begin{align}
E_k^{\text{EH,IC}}(\{\delta_E[n],\bold{q}_{m}[n]\})=\sum_{n=1}^N E_k(\delta_E[n],\{\bold{q}_{m}[n]\}).\label{havestedenergy}
\end{align}

Next, we consider the uplink WIT at time slot $n\in\mathcal N$ with duration $\delta_I[n]$. Each IoT-device $k\in\{1,2\}$ sends independent information signal $\chi_k[n]$ towards its associated UAV $k$, where $\chi_k[n]$'s are assumed to be independent and identically distributed (i.i.d.) circularly symmetric complex Gaussian (CSCG) random variables with zero mean and unit variance, i.e., $\chi_k[n]\sim\mathcal{CN}(0,1)$. Let $Q_k[n],k\in\{1,2\},n\in\mathcal N$, denote the transmit power of IoT-device $k$ at slot $n$. Accordingly, the signal-to-interference-plus-noise ratio (SINR) at the receiver of UAV $k\in\{1,2\}$ is
\begin{align}
\gamma_{k}(\{Q_i[n]\},\bold{q}_{k}[n])=\frac{Q_k[n]g_{k,k}[n]}{Q_{\bar k}[n]g_{\bar k,k}[n]+\sigma^2},
\end{align}
where $\sigma^2$ denotes the additive white Gaussian noise (AWGN) power at the information receiver of each UAV and $\bar k\in\{1,2\}\setminus\{k\}$. Accordingly, the achievable data-rate of IoT-device $k\in\{1,2\}$ at slot $n\in\mathcal N$ is given by
\begin{align}
r_k(\{Q_i[n]\},\delta_{I}[n],\bold{q}_{k}[n])=\delta_{I}[n]{\rm{log_2}}\left(1+\gamma_{k}(\{Q_i[n]\},\bold{q}_{k}[n])\right).
\end{align}

Therefore, the average data-rate throughput of IoT-device $k\in\{1,2\}$ over the entire period in bits-per-second-per-Hertz (bps/Hz) is given by
\begin{align}
R_k(\{Q_i[n],\delta_{I}[n],\bold{q}_k[n]\})=\frac{1}{T}\sum_{n=1}^N r_k(\{Q_i[n]\},\delta_{I}[n],\bold{q}_{k}[n]).
\end{align}
Note that for the purpose of exposition, we consider that the energy consumption of each IoT-device is dominated by the transmit power for uplink WIT. In this case, the total energy consumption at IoT-device $k\in\{1,2\}$ is
\begin{align}\label{energyconsumption}
E_k^{\text{TX,IC}}(\{Q_k[n],\delta_{I}[n]\}) = \sum_{n=1}^N Q_k[n]\delta_{I}[n].
\end{align}

In order to achieve sustainable operation for WPCN, we consider the energy neutrality constraint at each IoT-device $k$, such that the IoT-device's energy consumption for uplink WIT (i.e., $E_k^{\text{TX,IC}}(\{Q_k[n],\delta_{I}[n]\})$ in (\ref{energyconsumption})) does not exceed the energy harvested from the downlink WPT (i.e., $E_k^{\text{EH,IC}}(\{\delta_E[n],\bold{q}_m[n]\})$ in (\ref{havestedenergy})) in each period. As a result, we have
\begin{align}\label{energyconstraint}
\sum_{n=1}^N Q_k[n]\delta_{I}[n] \le \sum_{n=1}^N\sum_{m=1}^2\frac{ \eta P\beta_{\rm 0}\delta_E[n] }{\lVert\bold{q}_{m}[n]-\bold{w}_k\rVert^2+H^2}, \forall k\in\{1,2\}.
\end{align}

Our objective is to maximize the uplink common (or minimum) throughput of the two IoT-devices (i.e., $\min_{k\in\{1,2\}}R_k(\{Q_i[n],\delta_{I}[n],\bold{q}_k[n]\})$), subject to the UAVs' pre-determined initial and final location constraints in (\ref{IFC}), the  UAVs' maximum flying speed constraints in (\ref{speedcon}), the collision avoidance constraints in (\ref{conconcon}), and the IoT-devices' energy neutrality constraints in (\ref{energyconstraint}). The optimization variables include the UAVs' trajectories $\{\bold{q}_{m}[n]\}$, the time allocation $\{\delta_{I}[n],\delta_{E}[n]\}$, and the two IoT-devices' transmit power allocations $\{Q_k[n]\}$. As a result, the uplink common throughput maximization problem is formulated as
\begin{align}
\text{(P1)}&:\max_{\{\bold{q}_{m}[n],\delta_{I}[n],\delta_{E}[n],Q_k[n]\}}~ \min_{k\in\{1,2\}}R_k(\{Q_k[n],\delta_{I}[n],\bold{q}_m[n]\})\nonumber\\
&~~~~~\mathrm{s.t.}
~\delta_E[n]+\delta_{I}[n] \le \delta, \forall n\in\mathcal N\label{con14}\\
&~~~~~~~~~~\delta_{I}[n]\ge 0, \delta_E[n]\ge 0,\forall n\in\mathcal N\label{con16}\\
&~~~~~~~~~~\text{(\ref{IFC}), (\ref{speedcon}), (\ref{conconcon}), and (\ref{energyconstraint}).}\nonumber
\end{align}
Notice that for problem (P1), the objective function is non-concave, and constraints (\ref{speedcon}), (\ref{conconcon}), and (\ref{energyconstraint}) are all non-convex, due to the coupling between the time allocation ($\delta_{I}[n]$ and $\delta_E[n]$) and the trajectory ($\bold{q}_m[n]$) or transmit power ($Q_k[n]$). Therefore, (P1) is a non-convex optimization problem, which is difficult to be optimally solved in general. We will deal with problem (P1) in Section \ref{sec:P2}.

\vspace{-15pt}\subsection{Scenario with CoMP}
In this subsection, we consider the scenario with CoMP transmission/reception at the two UAVs, in which the two UAVs charge the ground IoT-devices via collaborative energy beamforming in the downlink\cite{SLee}, and forward their received signals from IoT-devices to a central controller (CC) for joint detection in the uplink. Similar to Section \ref{sec:inter}, at each time slot $n\in\mathcal N$, we assume that the WPT and WIT are operated over orthogonal sub-slots with duration $\rho_E[n]\ge 0$ and $\rho_I[n]\ge 0$, respectively, with $\rho_E[n]+\rho_I[n]\le \delta,\forall n\in\mathcal N.$

First, we consider the downlink WPT at time slot $n\in\mathcal N$ with duration $\rho_E[n]$, during which each UAV $m\in\{1,2\}$ uses a constant power $P$ to transmit a common energy signal $s^{\text{E}}$ (with normalized power $\mathbb{E}[|s^{\text{E}}|^2] = 1$)  with properly designed phase $\phi_m[n]$, i.e., $\sqrt{P}e^{j\phi_m[n]}s^{\text{E}}$, for charging the two IoT-devices. For simplicity, we design the collaborative energy beamforming based on the TDMA principle, in which both UAVs cooperatively design their signal phases to form energy beamforming vectors towards each of the two IoT-devices sequentially in a TDMA manner. Hence, we further divide the sub-slot for WPT with duration $\rho_E[n]$ into two sub-sub-slots each with duration $\rho_{E,k}[n]$ for IoT-device $k\in\{1,2\}$, where $\sum_{k=1}^2 \rho_{E,k}[n]\le \rho_E[n]$. In sub-sub-slot $k\in\{1,2\}$, the transmitted signal at each UAV $m\in\{1,2\}$ is given as $\sqrt{P}e^{j\phi_{m,k}[{n}]}s^{\text{E}}$, where $\phi_{m,k}[{n}]=-\theta_{k,m}[n]$ with $\theta_{k,m}[n]$ given in (\ref{channel}) denoting the phase of the channel coefficient from UAV $m$ to IoT-device $k$ at time slot $n$, such that the energy signals from both UAVs will be coherently combined at IoT-device $k$. Note that as the channel phases change in a much faster time scale (e.g., on a millisecond level) than the duration of slot or even sub-sub-slot, the UAVs should track such changes to design the transmit energy beamforming (via, e.g., efficient channel estimation).
Therefore, in the $k$-th sub-sub-slot, the received power by IoT-device $k$ (with coherent combining) is given by
\begin{align}
\bar{E}_k(\{\bold{q}_m[n]\})=\eta \mathbb{E}\left[\left|\sum_{m=1}^2 \sqrt{P}|h_{k,m}[n]| s^E\right|^2\right] = \eta P\left(\sum_{m=1}^2\sqrt{ \frac{\beta_0}{\lVert{\bold{q}}_m[n]-\bold{w}_k\rVert^2+H^2}}\right)^2,
\end{align}
and that by the other IoT-device $\bar k\in\{1,2\}\setminus\{k\}$ (without coherent combining) is
\begin{align}
{E}'_{\bar k}(\{\bold{q}_m[n]\})=\eta \mathbb{E}\left[\left|\left(\sum_{m=1}^2 \sqrt{P}h_{\bar k,m}[n] e^{j (\theta_{k ,m}[n] - \theta_{\bar k,m}[n])}\right) s^E\right|^2\right] = \eta P\sum_{m=1}^2 \frac{\beta_0}{\lVert{\bold{q}}_m[n]-\bold{w}_{\bar k}\rVert^2+H^2},
\end{align}
where the expectation is taken with respect to the randomness in the channel phases and the energy signal $s^{\text{E}}$. Therefore, the total energy harvested by IoT-device $k$ is given by
\begin{align}
{E}_{k}^{\text{EH,CoMP}}(\{\rho_{E,k}[n],\bold{q}_m[n]\})=&\sum_{n=1}^{{N}}\left(\rho_{E,k}[n]\bar{E}_k(\{\bold{q}_m[n]\})+\rho_{E,\bar k}[n]{E}'_{k}(\{\bold{q}_m[n]\})\right).\label{equ99}
\end{align}

Next, we consider the WIT at time slot $n\in\mathcal N$ with duration $\rho_I[{n}]$, in which both UAVs jointly detect the simultaneously transmitted signals by the two ground IoT-devices. In this case, the uplink WIT forms a virtual $2\times2$ multiple-input multiple-output (MIMO) system, for which the received signal $\mv{s}[{n}]=[s_1[n],s_2[n]]^T\in \mathbb{C}^{2\times 1}$ at the two UAVs is given by
\begin{align}
\mv{s}[{n}]=\sum_{k=1}^2 \mv{h}_k[{n}]\sqrt{Q_k[{n}]}\chi_k[n]+\mv{z}[n] =\mv{H}[{n}]\boldsymbol{\chi}[n]+\mv{z}[n],
\end{align}
where $\boldsymbol{\chi}[n]=[\sqrt{Q_1[{n}]}\chi_1[n],\sqrt{Q_2[{n}]}\chi_2[n]]^T\in \mathbb{C}^{2\times 1}$ denotes the transmitted signals by the two ground IoT-devices, $\mv{h}_k[{n}]=[h_{k,1}({\bold{q}}_1[{n}]), h_{k,2}({\bold{q}}_2[{n}])]^T$ denotes the channel vector from IoT-device $k$ to the two UAVs, $\mv{H}[{n}]=[\mv{h}_1[{n}],\mv{h}_2[{n}]]$, and $\mv{z}[n]=[z_1[n], z_2[n]]^T$ with $z_k[n]\sim\mathcal{CN}(0,\sigma^2)$ denoting the AWGN at the receiver of UAV $k$.

In this case, by using the linear receive beamforming, the processed signal at the two UAVs becomes
\begin{align}
\tilde{\mv{s}}=\mv{W}[{n}]\mv{s}[{n}] =\mv{W}[{n}]\mv{H}[{n}]\boldsymbol{\chi}[n]+\mv{W}[{n}]\mv{z}[n],
\end{align}
where $\mv{W}[{n}]=[\mv{w}_1[{n}],\mv{w}_2[{n}]]$, and $\mv{w}_k[{n}]\in\mathbb{C}^{2\times1}$ denotes the beamforming vector for detecting the signal $\chi_k[n]$ sent from IoT-device $k$, with $\|\mv{w}_k[{n}]\|=1$.

We adopt the zero-forcing (ZF) beamforming to eliminate the inter-device interference, such that $\mv{w}_k[{n}]^H\mv{h}_{\bar k}[{n}]=0,\forall k\in\{1,2\}$ must hold.\footnote{Notice that in general, the minimum mean squared error (MMSE)-based beamforming is the optimal linear receive beamforming design for maximizing both the SINR and rate. The MMSE beamforming, however, will lead to complicated SINR and rate functions with respect to the UAV trajectory. To avoid this drawback for facilitating the trajectory design, we consider the ZF receive beamforming in this paper, which asymptomatically approaches to the optimal linear MMSE receiver for interference-limited scenario.}
As $\mv{H}[n]$ is a $2 \times 2$ square matrix that is non-singular with probability 1, we have $\mv{W}[n]=\mv{H}[n]^{-1}$. Accordingly, the received signal-to-noise-ratio (SNR) at the two UAVs for detecting the signal sent by IoT-device $k$ at slot ${n}\in{\mathcal{N}}$ is given by
\begin{align}
\tilde\gamma_k(Q_k[n],\bold{q}_k[n])=&\frac{Q_k[{n}] |\mv{w}_k[{n}]^H\mv{h}_{k}[{n}]|^2}{\sigma^2},\forall k\in\{1,2\}.
\end{align}
Notice that $\tilde\gamma_k(Q_k[n],\bold{q}_k[n])$ is a random variable due to the random channel phases. Therefore, we are interested in the average rate for each IoT-device $k\in \{1,2\}$ at slot $n\in\mathcal N$, which is given by
\begin{align}
\tilde{r}_k(Q_k[n],\rho_I[{n}],\bold{q}_k[n])&=\rho_I[{n}]\mathbb{E}\left[\log_2\left(1+\frac{Q_k[{n}] |\mv{w}_k[{n}]^H\mv{h}_{k}[{n}]|^2}{\sigma^2}\right)\right],\label{rate21}
\end{align}
where the expectation is taken with respect to the randomness in the channel phases.

As a result, the average data-rate throughput of IoT-device $k$ over the entire mission period in bps/Hz is given by
\begin{align}
\tilde{R}_k(\{Q_k[n],\rho_I[{n}],\bold{q}_k[n]\})=\frac{1}{T}\sum_{{n}=1}^{{N}}\tilde{r}_k(Q_k[n],\rho_I[{n}],\bold{q}_k[n]).
\end{align}

To guarantee the sustainable operation for the WPCN, we have the following energy neutrality constraint at each IoT-device $k$, which is similar to (\ref{energyconstraint}).
\begin{align}
\sum_{{n}=1}^{{N}}\rho_I[{n}]Q_k[{n}]\le {E}_{k}^{\text{EH,CoMP}}(\{\rho_{E,k}[n],\bold{q}_m[n]\}),\forall k\in\{1,2\}.\label{con69}
\end{align}

Accordingly, we formulate the uplink common throughput maximization problem for the CoMP scenario as
\begin{align}
&\text{(P2):}\max_{\{{Q}_k[{n}],\bold{q}_m[{n}],\rho_{E,k}[{n}], \rho_I[{n}]\}}~\min_{k\in\{1,2\}}~\tilde{R}_k(\{Q_k[n],\rho_I[{n}],\bold{q}_k[n]\})\nonumber\\
&~~~~~~~~\mathrm{s.t.}~\sum_{k=1}^2\rho_{E,k}[{n}]+\rho_I[{n}]\le\delta,\forall {n}\in{\mathcal N}\label{P5c2}\\
&~~~~~~~~~~~~~\rho_{E,k}[{n}]\ge0,\rho_I[{n}]\ge0,\forall k\in\{1,2\},{n}\in{\mathcal N}\label{P5c3}\\
&~~~~~~~~~~~~~\text{(\ref{IFC}), (\ref{speedcon}), (\ref{conconcon}), and (\ref{con69}).}\nonumber
\end{align}
Note that problem (P2) is also a non-convex optimization problem, which is even more challenging than problem (P1), due to the involvement of complicated expectation operations in the objective function. We will solve this problem in Section \ref{sec6}.

\vspace{-15pt}\section{Proposed Solution to Problem (P1) with Interference Coordination}\label{sec:P2}
In this section, we consider problem (P1) for the scenario with interference coordination. Before proceeding to solve (P1) for the general case, in Section \ref{subsec1} we first consider the special case with $T \rightarrow \infty$ to draw key insights. Then, in Section \ref{sec:V} we propose an efficient algorithm to obtain an effective solution to (P1) with finite $T$.
\vspace{-15pt}\subsection{Obtained Solution to (P1) with $T \rightarrow \infty$}\label{subsec1}
First, we consider the case when the UAV mission duration $T$ is sufficiently large (i.e., $T\to\infty$), such that the UAVs' flying time from one location to another is negligible. Therefore, the UAVs' maximum flying speed in (\ref{speedcon}) and the initial and final locations constraints in (\ref{IFC}) can be ignored. This is due to the fact that in this case, the UAVs can reach any locations (e.g., optimized hovering locations or initial/final locations) instantly  with the flight duration among these locations being finite and thus negligible. In this case, by introducing an auxiliary variable $R$, the uplink common throughput maximization problem (P1) is re-expressed as
\begin{align}
\text{(P1.1)}:&\max_{\{{Q}_k[{n}],\bold{q}_m[{n}],\rho_{E,k}[{n}], \rho_I[{n}]\},R}~ R\nonumber\\
\mathrm{s.t.}&~\frac{1}{T}\sum_{n=1}^N\delta_I[n]\log_2\left(1+\frac{Q_k[n]g_{k,k}[n]}{Q_{\bar k}[n]g_{\bar k,k}[n]+\sigma^2}\right)\ge R,\forall k\in\{1,2\}\\
&~\text{(\ref{conconcon}), (\ref{energyconstraint}), (\ref{con14}), and (\ref{con16}).}\nonumber
\end{align}
Notice that, without the constraints in (\ref{IFC}) and (\ref{speedcon}), the optimality of problem (P1.1) only depends on the UAVs' locations and durations (and the corresponding wireless resource allocations) over time, irrespective of the exact locations at each particular time instant. In other words, for UAV trajectories, as long as the UAVs stay at the same locations with same durations, the same objective value of (P1) can be achieved.\footnote{Consider two UAV trajectories $\{\bar {\bold{q}}_m(t)\}$ and $\{\hat {\bold{q}}_m(t)\}$. Suppose that the two UAVs stay at same locations with same duration $\bar \tau$, but at different time, e.g., $\bar {\bold{q}}_m(\bar t + t) = \hat {\bold{q}}_m(\hat t +t), \forall t\in [0,\bar \tau], \bar t \neq \hat t$. It is evident that the achieved objective by trajectory  $\{\bar {\bold{q}}_m(t)\}$ over period $[\bar t, \bar t + \bar \tau]$ is same as that by trajectory $\{\hat {\bold{q}}_m(t)\}$ over a different period $[\hat  t, \hat  t + \bar \tau]$ with same duration $\bar \tau$.} Based on this observation, let's denote the total durations for WPT and WIT by $\tau_E=\sum_{n=1}^N\delta_E[n]$ and $\tau_I=T-\tau_E=\sum_{n=1}^N\delta_I[n]$, respectively. Suppose that the WPT is operated during time interval $(0, \tau_E]$ and the WIT is operated over interval $(\tau_E, T]$. Then we can re-express problem (P1.1) as the following problem (P1.2) by replacing the summations in rate and energy terms as integrations, and changing slot index $n$ as continuous time instant $t$.
\begin{align}
\text{(P1.2)}:&\max_{\{{Q}_k(t),\bold{q}_m(t)\},0\le\tau_{E}\le T,R}~ R\nonumber\\
\mathrm{s.t.}&~~~\frac{1}{T}\int_{\tau_E}^T{\rm{log_2}}\left(1+\frac{Q_k(t)g_{k,k}(t)}{Q_{\bar k}(t)g_{\bar k,k}(t)+\sigma^2}\right)\text{d}t\ge R,\forall k\in\{1,2\}\\
~&~~~\int_{\tau_E}^T Q_k(t)\text{d}t\le \int_{0}^{\tau_E}\sum_{m=1}^2\frac{ \eta P\beta_{\rm 0} }{\lVert\bold{q}_{m}(t)-\bold{w}_k\rVert^2+H^2}\text{d}t,\forall k\in\{1,2\}\\
~&~~~\|\bold{q}_1(t)-\bold{q}_{2}(t)\|^2\ge d^2_{\text{min}},\forall t\in\mathcal T.
\end{align}

In the following, we solve problem (P1.2) by first optimizing over $\{{Q}_k(t)\}$, $\{\bold{q}_m(t)\}$, and $R$, under any given $\tau_E \in [0,T]$, and then adopt a 1D search to find the optimal $\tau_E$. First, we solve problem (P1.2) under any given $\tau_E \in [0,T]$, which can be decomposed into two subproblems for downlink WPT and uplink WIT, respectively. In particular, it can be shown that at the optimal solution, the amount of harvested energy by both IoT-devices should be equally maximized, for which we have the optimization problem for WPT for the interval $(0,\tau_E]$ as
\begin{align}
\max_{\|{\bold{q}}_1(t)-{\bold{q}}_{2}(t)\|^2\ge d^2_{\text{min}}}&~ \min_{k\in\{1,2\}} \int_{0}^{\tau_E}\sum_{m=1}^2\frac{ \eta P\beta_{\rm 0} }{\lVert{\bold{q}}_{m}(t)-\bold{w}_k\rVert^2+H^2}\text{d}t.\label{P1.1.1}
\end{align}
Let $E^{\text{EH,IC}*}(\tau_E)$ denote the objective value achieved by the obtained solution to problem (\ref{P1.1.1}), which corresponds to the harvested energy by each IoT-device during WPT under a given duration $\tau_E$.

Similarly, under a given $\tau_E$ and $E^{\text{EH,IC}*}(\tau_E)$, the communication rate of the two IoT-devices should also be equally maximized at the optimality of (P1.2). Hence, we have the optimization problem for WIT for the interval $[\tau_E,T)$ as
\begin{align}
&\max_{\{Q_k(t),\bold{q}_m(t)\},R}~ R\nonumber\\
\mathrm{s.t.}&~~~\frac{1}{T}\int_{\tau_E}^T{\rm{log_2}}\left(1+\frac{\frac{ Q_k(t)\beta_0}{\| {\bold{q}}_k(t)-\bold{w}_k\|^2+H^2}}{\frac{ Q_{\bar k}(t)\beta_0}{\| {\bold{q}}_k(t)-\bold{w}_{\bar k}\|^2+H^2}+\sigma^2}\right)\text{d}t\ge R,\forall k\in\{1,2\}\nonumber\\
~&~~~\int_{\tau_E}^T Q_k(t)\text{d}t\le E^{\text{EH,IC}*}(\tau_E),\forall k\in\{1,2\}\nonumber\\
~&~~~\|\bold{q}_1(t)-\bold{q}_{2}(t)\|^2\ge d^2_{\text{min}},\forall t\in(\tau_E,T].\label{P1.1.3}
\end{align}

Next, we solve problems (\ref{P1.1.1}) and (\ref{P1.1.3}) for WPT and WIT, respectively, under any given $\tau_E$.
\subsubsection{Solving Problem (\ref{P1.1.1}) for Common Energy Maximization under Given $\tau_E$}\label{sec:EH}
For problem (29), we consider that the two UAVs hover at fixed symmetric locations to equally maximize the harvested energy by the two IoT-devices. Suppose  that UAV 1 and UAV 2 stay at $\bold{q}_1(t)=\bar{\bold{q}}^E_1=({x}^E,0)$ and $\bold{q}_2(t)=\bar{\bold{q}}^E_2=(-{x}^E,0),$ respectively, $\forall t\in(0,\tau_E]$.{\footnote{It can be shown that in the case without collision avoidance constraint(i.e., $d_{\min} = 0$), the solution with UAVs hovering above the line connecting the two IoT-devices is optimal for common energy maximization for both IoT-devices. For the general case with $d_{\min} > 0$, we have conducted extensive simulations, which show that such a symmetric design is generally superior to other designs with unsymmetric UAV locations.}} As a result, solving problem (\ref{P1.1.1}) is equivalent to optimizing $x^E\ge 0$ to maximize the harvested energy by each IoT-device. With some simple manipulation, the optimization of $x^E$ becomes
\begin{align}
\max_{|{x}^E|\ge d_{\min}/2}&~\tau_E\eta P\beta_0\left(\frac{1}{({x}^E+D/2)^2+H^2}+\frac{1}{({x}^E-D/2)^2+H^2}\right).\label{P31}
\end{align}
By solving problem (\ref{P31}) via checking the first-order derivative of its objective function, we obtain the solution to problem (\ref{P1.1.1}) as follows.
\begin{lemma}\label{lem3.1}
At the obtained solution to problem (\ref{P1.1.1}), the two UAVs hover at two symmetric locations, denoted as $\{\bar{\bold{q}}^{E*}_m\}$, for downlink WPT, which are obtained by separately considering the following two cases.
\begin{itemize}
\item If IoT-devices' distance $D$ is no larger than $H/\sqrt{3}$, the optimal hovering locations of the two UAVs are given as ${\bold{q}}^{E}_1(t)=\bar{\bold{q}}^{E*}_1=(-d_{\text{min}}/2,0)$ and ${\bold{q}}^{E}_2(t)=\bar{\bold{q}}^{E*}_2=(d_{\text{min}}/2,0),\forall t\in(0,\tau_E]$;
\item If $D>2H/\sqrt{3}$, the optimal hovering locations of the two UAVs are
 given as ${\bold{q}}^{E}_1(t)=\bar{\bold{q}}^{E*}_1=(-\epsilon,0)$ and ${\bold{q}}^{E}_2(t)=\bar{\bold{q}}^{E*}_2=(\epsilon,0),\forall t\in(0,\tau_E]$, where
 \begin{align}
 \epsilon\triangleq\max\left(\sqrt{-(D^2/4+H^2)+\sqrt{D^4/4+H^2D^2}},d_{\text{min}}/2\right)\le \max\left(d_{\min}/2,D/2\right).\nonumber \end{align}
 \end{itemize}
\end{lemma}
\begin{IEEEproof}
	See Appendix \ref{AppA}.
\end{IEEEproof}

From Lemma \ref{lem3.1}, it is evident when the distance between the two ground IoT-devices is short, both UAVs should hover as close to the center of both IoT-devices as possible, while guaranteeing the minimum safety distance $d_{\min}$; otherwise, the two UAVs should hover at two symmetric locations between IoT-devices for efficient WPT. By substituting the optimal hovering locations $\{{\bold{q}}^{E}_m(t)\}$ into problem (\ref{P1.1.1}), the amount of harvested energy by both IoT-devices are equal, given by
\begin{align}
E_k^{\text{EH,IC}*}=E^{\text{EH,IC}*}(\tau_E)=\sum_{m=1}^2\frac{ \tau_E\eta P\beta_{\rm 0} }{\lVert{\bar{\bold{q}}}^{E*}_{m}-\bold{w}_k\rVert^2+H^2},\forall k\in\{1,2\}.\label{Estar}
\end{align}

\subsubsection{Solving Problem (\ref{P1.1.3}) for Common Rate Maximization under Given $\tau_E$}\label{UCTMP}
Next, we consider problem (\ref{P1.1.3}) under given $\tau_E$ and $E^{\text{EH,IC}*}(\tau_E)$ in (\ref{Estar}). For WIT over the time interval $(\tau_E,T]$, we consider the case that the two UAVs hover at fixed locations, i.e., $\bold{q}_m(t) = \bar{\bold{q}}_m^I, \forall m\in\{1,2\},  t\in (\tau_E,T]$. Furthermore, we consider a binary power control, in which the two IoT-devices either transmit simultaneously, or transmit in a TDMA manner with full power. This is motivated by the result that such binary power control is optimal for sum-rate maximization in a two-user interference channel with individual peak power constraints\cite{Binary}, which is thus an efficient choice under our setup with symmetric IoT-device and UAV locations. Accordingly, we have two transmission modes and obtain the solution to problem (\ref{P1.1.3}) as follows.
\begin{proposition}\label{proposition:3.1}
Under each of the two transmission modes, the obtained solution to problem (\ref{P1.1.3}) is given as follows.
\begin{itemize}
\item Mode 1 with simultaneous transmission: The two UAVs should hover at two symmetric hovering locations, denoted as  $\bold{q}_m^I(t)=\bar{\bold{q}}_m^{I*}=(\bar{x}_m^{I*},0),\forall m\in\{1,2\}, t\in(\tau_E,T]$ and each IoT-device transmits with power $Q_k(t)=Q^{(1)}=E^{\text{EH,IC}*}(\tau_E)/(T-\tau_E)$. Here, $\bar{x}_m^{I*}$'s are obtained based on the following equation
\begin{align}
\frac{(\bar{x}_1^{I*}+D/2)((\bar{x}_1^{I*}-D/2)^2+H^2)^2}{\beta_0Q^{(1)}/\sigma^2} -D(H^2+(D/2)^2-(\bar{x}_1^{I*})^2)=0,\label{xinmin}
\end{align}
 with $\max(D/2,d_{\min}/2)\le|\bar{x}_1^{I*}|\le\max(d_{\min}/2,\sqrt{(D/2)^2+H^2})$. Accordingly, the achievable common data-rate is $R_k=R_{\text{ST}}(\tau_E)=\frac{T-\tau_E}{T} {\rm{log_2}}\left(1+\frac{\frac{Q^{(1)}\beta_0}{({x}^{I*}_k-x_k)^2+H^2}}{\frac{ Q^{(1)}\beta_0}{({x}^{I*}_k-x_{\bar k})^2+H^2}+\sigma^2}\right),\forall k\in\{1,2\}$.
\item Mode 2 with TDMA transmission: Each UAV transmits over a half portion over the time interval $(\tau_E, T]$, and each UAV should hover exactly above its corresponding IoT-device during transmission, with $\bold{q}_1^I(t) = \bold{w}_1,\forall t\in(\tau_E,T/2+\tau_E/2]$, $\bold{q}_2^I(t) = \bold{w}_2, \forall t\in(T/2+\tau_E/2, T],$ and each IoT-device transmits with power $Q_1(t)=Q^{(2)}=2E^{\text{EH,IC}*}(\tau_E)/(T-\tau_E), Q_2(t)=0,\forall t\in(\tau_E,T/2+\tau_E/2]$, and $Q_1(t)=0, Q_2(t)=Q^{(2)},\forall t\in(T/2+\tau_E/2,T]$. Accordingly, the achievable common data-rate is $R_k=R_{\text{TDMA}}(\tau_E)=\frac{T-\tau_E}{2T}\log_2 \left(1+\frac{Q^{(2)}\beta_0/\sigma^2}{H^2}\right),\forall k\in\{1,2\}$.
\end{itemize}
\end{proposition}
\begin{IEEEproof}
As the optimal solution to Mode 2 with TDMA transmission can be easily verified, we only consider Mode 1 with simultaneous transmission, in which the two IoT-devices send information to their corresponding UAVs simultaneously by using fixed (and full) transmit power $Q^{(1)}=E^{\text{EH,IC}*}/(T-\tau_E)$. Note that at the UAVs' optimal hovering locations, the uplink communication throughput of each IoT-devices must be equal. Specifically, the two UAVs hover at two fixed symmetric locations, denoted by $\bold{q}_1^I(t)=\bar{\bold{q}}_1^I=(\bar{x}_1^I,0)$ and $\bold{q}_2^I(t)=\bar{\bold{q}}_2^I=(-\bar{x}_1^I,0),\forall t\in(\tau_E,T]$, to maximize the uplink common throughput. As a result, we can equivalently transform problem (\ref{P1.1.3}) into a sum-rate maximization problem as follows, in which
we only need to optimize one UAV's hovering locations $\bar{x}_1^I$,
\begin{align}
	\max_{|\bar{x}^I_1|\ge d_{\min}/2}&~\frac{T-{\tau}_E}{T} {\rm{log_2}}\left(1+\frac{\frac{Q^{(1)}\beta_0}{(\bar{x}^I_1+D/2)^2+H^2}}{\frac{ Q^{(1)}\beta_0}{(\bar{x}^I_1-D/2)^2+H^2}+\sigma^2}\right).\label{P28}
\end{align}
By checking the first-order derivative of objective function in (\ref{P28}), we can obtain the optimal $\bar{x}_1^{I*}$ based on the equation in (\ref{xinmin}). With the obtained optimal solution, it can be shown that
\begin{align}
	\max(D/2,d_{\min}/2)\le|\bar{x}_1^{I*}|\le\max(d_{\min}/2,\sqrt{(D/2)^2+H^2}). \nonumber
\end{align}
Thus, we have obtained the UAVs' optimal hovering locations $\{\bar{\bold{q}}_m^{I*}\}=\{(\bar{x}_m^{I*},0)\}$. Accordingly, the achieved common data-rate can be obtained as $R_{\text{ST}}(\tau_E)$. Therefore, Proposition \ref{proposition:3.1} is proved.
\end{IEEEproof}

By comparing the two optimal values of $R_{\text{ST}}(\tau_E)$ and $R_{\text{TDMA}}(\tau_E)$ with the two transmission modes, the optimal transmission mode can be obtained.

\subsubsection{Complete Algorithm for Solving (P1.2)}
Finally, it remains to obtain the optimal $\tau_E^*$. This is attained by using a 1D exhaustive search over $(0,T]$ by comparing the correspondingly achievable data-rate throughput in Proposition \ref{proposition:3.1}. With $\tau_E^*$, the corresponding achievable data-rate throughput and hovering locations in Lemma \ref{lem3.1} and Proposition \ref{proposition:3.1} become the obtained solution to problem (P1.2) or (P1.1). Therefore, problem (P1.2) is solved, for which the following insights can be obtained.


\begin{remark}\label{Remark:3.2}
The obtained solution to problem (P1.2) (or (P1.1)) has a multi-location-hovering structure as follows. During downlink WPT period, the UAVs hover at two symmetric locations given in Lemma \ref{lem3.1} that are generally above the center between the two IoT-devices with $|\bar{\bold{q}}_m^{E*}|\le \max(D/2,d_{\min}/2),\forall m\in\{1,2\}$; while during uplink WIT period, the UAVs hover at a different set of hovering locations given in Proposition \ref{proposition:3.1}, which are far away from the two IoT-devices with $|\bar{\bold{q}}_m^{I*}|\ge D/2,\forall m\in\{1,2\}$. Furthermore, for uplink WIT, there generally exist two transmission modes with simultaneous transmission and TDMA transmission, in which the UAVs' hovering locations are different, depending on the channel gain between UAVs and IoT-devices.
\end{remark}
\begin{figure}
\setlength{\abovecaptionskip}{-2.mm}
\setlength{\belowcaptionskip}{-2.mm}
\centering
    \includegraphics[width=8cm]{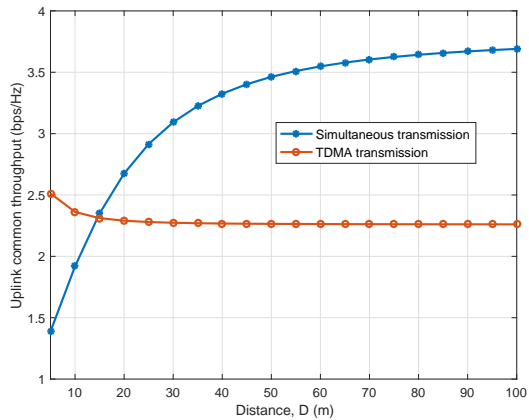}
\caption{The uplink common throughput versus the IoT-devices' distance $D$ in the interference coordination scenario with $H=5$~m.} \label{fig:10}
\end{figure}

{\it Example:} For illustration, Fig. \ref{fig:10} shows the uplink common throughput versus the distance $D$ between the two IoT-devices, where the same parameters as those in Section \ref{sec9} are used. It is observed that when $D$ is large, Mode 1 with simultaneous transmission outperforms Mode 2 with TDMA transmission, whereas when $D$ is small, the opposite is true. This is expected, as in the former case, the two UAVs can enjoy the benefit of spectrum reuse with weak co-channel interference; while in the later case, TDMA is preferred since it can avoid the strong interference. Moreover, the uplink common throughput achieved by TDMA remains nearly unchanged when $D$ becomes sufficiently large, which is due to the fact that the wireless energy transferred through the cross channel and the co-channel interference become negligible in this case.

\vspace{-15pt}\subsection{Proposed Solution to Problem (P1) with Finite $T$}\label{sec:V}
In this subsection, we propose an alternating optimization based solution to problem (P1) for the general case with finite $T$.
In the following, we optimize the time allocation $\{\delta_I[n],\delta_E[n]\}$, the IoT-devices' transmit power $\{Q_k[n]\}$, and the UAVs' trajectories $\{\bold{q}_m[n]\}$, sequentially in an alternating manner, with all other variables given.

\subsubsection{Time Allocation}
For any given UAVs' trajectories $\{\bold{q}_m[n]\}$ and IoT-device transmit power $\{Q_k[n]\}$, the time allocation problem in (P1) can be written as
\begin{align} \text{(P1-Time):}&\max_{\{\delta_I[n],\delta_E[n]\}}~\min_{k\in\{1,2\}}R_k(\{Q_k[n],\delta_{I}[n],\bold{q}_m[n]\})~~\mathrm{s.t.}~~\text{(\ref{energyconstraint}), (\ref{con14}), and (\ref{con16})}\nonumber,
\end{align}
which is a linear program (LP), and can be optimally and efficiently solved by standard convex optimization techniques, such as the interior point method\cite{Boyd2004}.

\subsubsection{Transmit Power Allocation}
For any given time allocation $\{\delta_I[n],\delta_E[n]\}$ and UAVs' trajectories $\{\bold{q}_m[n]\}$, the transmit power allocation problem in (P1) can be written as
\begin{align} \text{(P1-Power):}&\max_{\{Q_k[n]\}}~~\min_{k\in\{1,2\}}R_k(\{Q_k[n],\delta_{I}[n],\bold{q}_m[n]\})~~\mathrm{s.t.}~\text{(\ref{energyconstraint})}\nonumber,
\end{align}
which is a non-convex optimization problem due to the non-concave objective function $r_k(\{Q_k[n]\},\delta_{I}[n],\bold{q}_{k}[n])$ with respect to $\{Q_k[n]\}$. To tackle the non-concave objective function, we use the SCA to update the IoT-devices' transmit power $\{Q_k[n]\}$ in an iterative manner by approximating the transmit power allocation problem into a convex problem at each iteration. Note that any convex function is globally lower-bounded by it first-order Taylor expansion at any point. By denoting $\{Q_k^{(l)}[n]\}$ as the transmit power of IoT-device $k$ in the $l$-th iteration, we have
\begin{align}
&r_k(\{Q_k[n]\},\delta_{I}[n],\bold{q}_{k}[n])\ge\delta_{I}[n]{\rm{log_2}}\left(\sum_{i=1}^{2}Q_{i}[ n]g_{i,k}[n]+\sigma^2\right)\nonumber\\
&-\delta_{I}[n]\log_2\left(Q_{\bar k}^{(l)}[n]g_{\bar k,k}[n] +\sigma^2\right)-\frac{\delta_{I}[n]g_{\bar k,k}[n]\log_2(e)}{Q_{\bar k}^{(l)}[n]g_{\bar k,k}[n] +\sigma^2}(Q_{\bar k}[n]-Q_{\bar k}^{(l)}[n])\nonumber\\
&\triangleq r_k^{\text{low}-(l)}(\{Q_k[n]\},\delta_{I}[n],\bold{q}_{k}[n]).\label{upper34}
\end{align}
Therefore, in each iteration $l$ with given local point $\{Q_{k}^{(l)}[n]\}$, we replace $r_k(\{Q_k[n]\},\delta_{I}[n],\bold{q}_{k}[n])$ as its lower bound $r_k^{\text{low}-(l)}(\{Q_k[n]\},\delta_{I}[n],\bold{q}_{k}[n])$ in (\ref{upper34}). Accordingly, the transmit power allocation problem (P1-Power) is transformed into a convex optimization problem, which can be efficiently solved by standard convex optimization solvers such as CVX\cite{CVX}. It follows that after each iteration $l\ge1$, the objective value of problem (P1-Power) achieved by $\{Q_{k}^{(l)}[n]\}$ is at least no smaller than that achieved by $\{Q_{k}^{(l-1)}[n]\}$ in the previous iteration $(l-1)$. As the optimal value of problem (P1-Power) is bounded, the SCA-based algorithm will lead to a converged solution.

\subsubsection{UAVs' Trajectories Optimization}\label{traopt}
For any given time allocation $\{\delta_I[n],\delta_E[n]\}$ and IoT-devices' transmit power $\{Q_k[n]\}$, the UAVs' trajectories optimization problem in (P1) can be written as
\begin{align}
\text{(P1-Tra):}&\max_{\{\bold{q}_m[n]\}}~\min_{k\in\{1,2\}}R_k(\{Q_k[n],\delta_{I}[n],\bold{q}_m[n]\})\nonumber\\
&\mathrm{s.t.}~\text{(\ref{IFC}), (\ref{speedcon}), (\ref{conconcon}), and (\ref{energyconstraint})}\nonumber,
\end{align}
which is non-convex due to the non-concave objective function and non-convex constraints in (\ref{conconcon}) and (\ref{energyconstraint}). Similar to the transmit power allocation problem, we use the SCA to update the UAVs' trajectories $\{\bold{q}_m[n]\}$ in an iterative manner by approximating the UAVs' trajectories optimization problem into a convex optimization problem at each iteration. We denote $\bold{q}_{m}^{(l)}[n]$ as the trajectory of UAV $m$ in the $l$-th iteration. As any convex function is globally lower-bounded by its first-order Taylor expansion at any point, we have

\vspace{-1em}\begin{footnotesize}
\begin{align}
&r_k(\{Q_k[n]\},\delta_{I}[n],\bold{q}_{k}[n])\nonumber\\
&\ge\delta_{I}[n]\Bigg(\hat r_k^{(l)}[n]-\sum_{i=1}^{2}\frac{\frac{Q_{i}[ n]\beta_0}{(\lVert\bold{q}^{(l)}_k[ n] -\bold{w}_{i}\rVert^2+H^2)^2}\log_2(e)}{\sum_{z=1}^{2}\frac{Q_{z}[ n]\beta_0}{\lVert\bold{q}^{(l)}_k[n] -\bold{w}_{z}\rVert^2+H^2}+\sigma^2}\left(\lVert\bold{q}_k[n] -\bold{w}_{i}\rVert^2-\lVert\bold{q}_k^{(l)}[n] -\bold{w}_{i}\rVert^2\right)-\bar{r}_k[n]\Bigg)\nonumber\\
&\triangleq r_k^{\text{low}-(l)}(\{Q_k[n]\},\delta_{I}[n],\bold{q}_{k}[n]) ,\label{lbr}\\
&E_k^{\text{EH,IC}}(\{\delta_E[n],\bold{q}_{m}[n]\})\nonumber\\
&\ge\sum_{m=1}^2\left(\frac{2\eta P\beta_{\rm 0}\delta_E[n]}{H^2+\lVert\bold{q}_{m}^{(l)}[n]-\bold{w}_k\rVert^2}- \frac{\eta P\beta_{\rm 0}\delta_E[n](H^2+\lVert\bold{q}_{m}[n]-\bold{w}_k\rVert^2)} {(H^2+\lVert\bold{q}_{m}^{(l)}[n]-\bold{w}_k\rVert^2)^2}\right)\nonumber\\
&\triangleq E_k^{\text{low}-(l)}(\{\delta_E[n],\bold{q}_{m}[n]\}),\label{lbe}\\
&\|\bold{q}_1[n]-\bold{q}_{2}[n]\|^2\ge-\|\bold{q}_{1}^{(l)}[ n]-\bold{q}_{2}^{(l)}[ n]\|^2 +2(\bold{q}_1^{(l)}[ n]-\bold{q}_{2}^{(l)}[ n])^T(\bold{q}_1[ n]-\bold{q}_{2}[ n]),\label{lbq}
\end{align}
\end{footnotesize}where $\hat r_k^{(l)}[n]={\rm{log_2}}\left(\sum_{i=1}^{2}Q_{i}[ n]g_{i,k}(\bold{q}_k^{(l)}[n])+\sigma^2\right)$ and $\bar r_k[n]=\log_2\left(Q_{\bar k}[n]g_{\bar k, k}(\bold{q}_k[n])+\sigma^2\right)$. Based on the approximations in (\ref{lbr}), (\ref{lbe}), and (\ref{lbq}), problem (P1-Tra) can be transformed into a convex problem in each iteration, which can be solved by CVX.

Note that as problem (P1) is a non-convex optimization problem that generally possesses multiple locally optimal solutions, the performance of the SCA-based iterative algorithm critically depends on the chosen initial point for iteration. In this case, we choose the SHF trajectory as the initial point. In the SHF trajectory, the UAVs fly at the maximum speed from initial locations to the optimal hovering locations for downlink WPT $\{\bar{\bold{q}}^{E*}_m\}$, then fly to the optimal hovering locations for uplink WIT $\{\bar{\bold{q}}^{I*}_m\}$ and finally fly to final locations. Suppose that the minimum flying duration for UAVs to visit these hovering locations at the maximum speed is $T_{\text{fly}}$. If the total duration $T<T_{\text{fly}}$, then we consider the direct flight trajectory, i.e., each of the two UAVs simply flies straightly from the initial location to the final location at a uniform flying speed. In this case, the time and power allocation problems can be solved similarly as those in problem (P1-Time) and (P1-Power), respectively. As a result, the alternating-optimization-based approach in this section can always achieve a common throughput no smaller than the SHF trajectory design.

In summary, we propose an algorithm for problem (P1) via solving three subproblems in an alternating manner. Notice that the above solutions to these subproblems converge to solutions with non-decreasing objective values. As the optimal value of (P1) is bounded from above, it is evident that our proposed design can result in a converged solution to (P1).


\vspace{-15pt}\section{Proposed Solution to Problem (P2) with CoMP}\label{sec6}
In this section, we consider problem (P2) for the scenario with CoMP transmission/reception at UAVs. Note that in this case, it is difficult to obtain an explicit function for the average communication rate at each time slot $n$ given in (\ref{rate21}) with respect to the UAVs' trajectories. To overcome this challenge, we utilize efficient bounding and approximation results in \cite{liuliang} to derive tractable approximate functions of $\tilde{r}_k[{n}]$'s.

\begin{lemma}
At each time slot ${n}$, $\tilde{r}_k[{n}]$ is bounded by
\begin{align}
\tilde{r}_k[{n}]\le\log_2\left(1+\frac{1}{2}\sum_{m=1}^2 \frac{Q_k[{n}]\beta_0/\sigma^2} {\lVert{\bold{q}}_m[{n}]-\bold{w}_k\rVert^2+H^2}\right),\forall k\in\{1,2\},n\in\mathcal{N}.\label{upper}
\end{align}
\end{lemma}

In this case, we approximate problem (P2) by replacing $\tilde{r}_k[{n}]$ with its upper bound in (\ref{upper}), and introducing an auxiliary variable $R_c$.
\begin{align}
&\text{(P2.1):}\max_{\{{Q}_k[{n}],\bold{q}_m[{n}],\rho_{E,k}[{n}], \rho_I[{n}]\},R_c}~R_c\nonumber\\
&\mathrm{s.t.}\frac{1}{T}\sum_{{n}=1}^{{N}} \rho_I[{n}]\log_2\left(1+\frac{1}{2}\sum_{m=1}^2 \frac{Q_k[{n}]\beta_0/\sigma^2} {\lVert{\bold{q}}_m[{n}]-\bold{w}_k\rVert^2+H^2}\right)\ge R_c,\forall k\in\{1,2\}\label{P2.1con}\\
&~~~~~\text{(\ref{IFC}), (\ref{speedcon}), (\ref{conconcon}), (\ref{con69}), (\ref{P5c2}), and (\ref{P5c3}).}\nonumber
\end{align}
Similar to problem (P1), in this section, we first consider the case when $T\to\infty$ to draw essential design insights, and then propose an efficient algorithm to solve (P2.1) for the general case with finite $T$.

\vspace{-15pt}\subsection{Obtained solution to (P2.1) with $T\to\infty$}
First, consider the case that the UAV mission duration $T$ is sufficiently long, such that we can ignore the UAVs' maximum flying speed constraint in (\ref{speedcon}) and the initial and final locations constraints in (\ref{IFC}). In this case, the uplink common throughput maximization problem (P2.1) is reformulated as
\begin{align}
&\text{(P2.2):}\max_{\{{Q}_k[{n}],\bold{q}_m[{n}],\rho_{E,k}[{n}], \rho_I[{n}]\},R_c}~R_c\nonumber\\
&~~~~~~~~~~~\mathrm{s.t.}~~\text{(\ref{conconcon}), (\ref{con69}), (\ref{P5c2}), (\ref{P5c3}), and (\ref{P2.1con}).}\nonumber
\end{align}


Similar to (P1.1), the optimization of problem (P2.2) only depends on the UAVs' locations and durations over time. Suppose that the total durations for WPT and WIT is denoted by $\tilde{\tau}_{E,k}=\sum_{n=1}^N\rho_{E,k}[n],\forall k\in\{1,2\}$ and $\tilde{\tau}_I=T-\sum_{k=1}^2\tilde{\tau}_{E,k}=\sum_{n=1}^N\rho_{I}[n]$, respectively. We have $\tilde{\tau}_{E,1} = \tilde{\tau}_{E,2} = \tilde{\tau}_E/2$ due to the symmetric nature of the two IoT-devices. Specifically, the collaborative energy beamforming towards IoT-device 1 and IoT-device 2 are operated during time intervals $(0,\tilde{\tau}_{E}/2]$ and $(\tilde{\tau}_{E}/2,\tilde{\tau}_{E}]$, and the WIT is operated over interval $(\tilde{\tau}_{E},T]$. Accordingly, similar to that for (P1.2), problem (P2.2) can be equivalently re-expressed as
\begin{align}
&\text{(P2.3):}\max_{\{{\bold{q}}_m(t),Q_k(t),0\le\tilde{\tau}_{E}\le T\},R_c}~R_c\nonumber\\
&\mathrm{s.t.}\frac{1}{T}\int_{\tilde{\tau}_{E}}^{T} \log_2\left(1+\frac{1}{2}\sum_{m=1}^2 \frac{Q_k(t)\beta_0/\sigma^2} {\lVert{\bold{q}}_m(t)-\bold{w}_k\rVert^2+H^2}\right)\text{d}t\ge R_c,\forall k\in\{1,2\}\\
&~~~\int_{\tilde{\tau}_{E}}^{T}Q_k(t)\text{d}t\le\eta P \left(\int_{0}^{\tilde{\tau}_{E}/2}\bar{E}_k(\{\bold{q}_m(t)\}) \text{d}t+\int_{\tilde{\tau}_{E}/2}^{\tilde{\tau}_{E}}{E}'_k(\{\bold{q}_m(t)\})\text{d}t\right)\\
&~~~\|\bold{q}_1(t)-\bold{q}_2(t)\|^2\ge d_{\min}^2,\forall t\in\mathcal T.
\end{align}
Similar to that in Section \ref{subsec1}, to obtain insightful solutions to problem (P2.3) or (P2.2), for any given durations $\tilde{\tau}_{E}$ and $\tilde{\tau}_I$, we first consider the following common energy maximization problem in the downlink WPT. As the charging duration for the two IoT-devices are equal, during $(0,\tilde {\tau}_E/2]$ and $(\tilde {\tau}_E/2,\tilde \tau_E]$, the hovering locations of both UAVs should be symmetric, which leads to the following problem.
\begin{align}
&\max_{\{{\bold{q}}_m(t)\}}\min_{k\in\{1,2\}}\eta P \left(\int_{0}^{\tilde{\tau}_{E}/2}\bar{E}_k(\{\bold{q}_m(t)\}) \text{d}t+\int_{\tilde{\tau}_{E}/2}^{\tilde{\tau}_{E}}{E}'_k(\{\bold{q}_m(t)\})\text{d}t\right)\nonumber\\
&\mathrm{s.t.}~\|{\bold{q}}_1(t)-{\bold{q}}_2(t)\|^2\ge d_{\min}^2,\forall t\in(0,\tilde{\tau}_{E}].\label{P58}
\end{align}
Let ${E}^{\text{EH,CoMP}*}(\tilde{\tau}_{E})$ denote the optimal value achieved by problem (\ref{P58}), which corresponds to the maximum harvested energy by each IoT-device. Accordingly, we have the common rate maximization problem in the uplink WIT as follows.
\begin{align}
\max_{\{{\bold{q}}_m(t),Q_k(t)\},R_c}&~R_c\nonumber\\
\mathrm{s.t.}&~\frac{1}{T} \int_{\tilde{\tau}_E}^T\log_2\left(1+\frac{1}{2}\sum_{m=1}^2 \frac{Q_k(t)\beta_0/\sigma^2} {\lVert{\bold{q}}_m(t)-\bold{w}_k\rVert^2+H^2}\right)\text{d}t\ge R_c\nonumber\\
&~\int_{\tilde{\tau}_E}^T Q_k(t)\text{d}t\le {E}^{\text{EH,CoMP}*}(\tilde{\tau}_{E}),\forall k\in\{1,2\}\nonumber\\
&~\|{\bold{q}}_1(t)-{\bold{q}}_2(t)\|^2\ge d^2_{\text{min}},\forall t\in(\tilde{\tau}_E,T].\label{P62}
\end{align}

In the following, we first obtain the UAV placement and power allocation solution to problems (\ref{P58}) and (\ref{P62}) for WPT and WIT, respectively, under any given $\tilde{\tau}_{E}$ and $\tilde{\tau}_I$, and then adopt a 1D exhaustive search to find the optimal time allocation of $\tilde{\tau}_{E}$, denoted by $\tilde{\tau}_{E}^*$.

\subsubsection{Solving Problem (\ref{P58}) for Common Energy Maximization under Given $\{\tilde{\tau}_{E}\}$}
It is evident that at the optimal solution to problem (\ref{P58}), the harvested energy by each IoT-device must be equal. Therefore, for the two intervals $(0, \tilde \tau_E/2]$ and  $(\tilde \tau_E/2, \tilde \tau_E]$, the two UAVs should stay at fixed hovering locations for efficient WPT, denoted by $\bold{q}_m^{E}(t)=\tilde{\bold{q}}_m^{E,1}=(\tilde{x}^{E,1}_m,0),\forall t\in(0,\tilde{\tau}_{E}/2],m\in\{1,2\}$, and $\bold{q}_m^{E}(t)=\tilde{\bold{q}}_m^{E,2}=(\tilde{x}^{E,2}_m,0),\forall t\in(\tilde{\tau}_{E}/2,\tilde{\tau}_{E}],m\in\{1,2\}$. Due to the symmetric nature, we have $\tilde{x}^{E,k}_1=-\tilde{x}^{E,\bar k}_2,\forall k\in\{1,2\}$. As a result, problem (\ref{P58}) is reduced to
\begin{align}
&\max_{\{|\tilde{x}_1^{E,k}-\tilde{x}_2^{E,k}|^2\ge d_{\min}^2\}}~\frac{\tilde{\tau}_{E} \eta P}{2}\sum_{k=1}^2~\left(\bar{E}_k(\{\tilde{\bold{q}}_m^{E,k}(t)\})+{E}'_k(\{\tilde{\bold{q}}_m^{E,k}(t)\})\right)\label{P63}.
\end{align}
Here, problem (\ref{P63}) is a non-convex optimization problem due to the non-concave objective function. In this case, we solve problem (\ref{P63}) via a 2D exhaustive search to obtain the UAVs' optimal hovering locations for downlink WPT. By substituting the optimal hovering locations $\{\tilde{\bold{q}}^{E,k*}_m=(\tilde{x}^{E,k*}_m,0)\}$, we have the common harvested energy of each IoT-device as ${E}_k^{\text{EH,CoMP}*}={E}^{\text{EH,CoMP}*}(\tilde{\tau}_{E}),\forall k\in\{1,2\}$.

\subsubsection{Solving Problem (\ref{P62}) for Common Rate Maximization under Given $\tilde{\tau}_{E}$}
For WIT, it can also be shown that at the optimal solution to problem (\ref{P62}), the average rates of both IoT-devices should be equal. In this case, UAV 1 should hover at $\bold{q}_1^I(t)=\tilde{\bold{q}}_2^I=(\tilde{x}^I,0)$ while UAV 2 should hover at the symmetric location $\bold{q}_2^I(t)=\tilde{\bold{q}}_2^I=(-\tilde{x}^I,0),\forall t\in(\tilde{\tau}_E,T]$. Note that since ZF receiver is used, both IoT-devices should transmit information to UAVs with the maximum power to achieve maximum data rate. Thus, we have $Q_1^{\text{CoMP}}=Q_2^{\text{CoMP}}={E}^{\text{EH,CoMP}*}(\tilde{\tau}_{E})/\tilde{\tau}_I$. Accordingly, problem (\ref{P62}) reduces to
\begin{align}
\max_{|\tilde{x}^I|\ge d_{\text{min}}/2}&~\sum_{k=1}^2~\frac{\tilde{\tau}_I}{T} \log_2\left(1+\frac{1}{2}\sum_{m=1}^2 \frac{Q_k^{\text{CoMP}}\beta_0/\sigma^2} {(\tilde{x}^I-x_k)^2+H^2}\right).\label{P62sum}
\end{align}
Hence, we can obtain the UAV hovering locations in closed-form, given in the following lemma.

\begin{lemma}\label{lem4.1}
The two UAVs' optimal hovering locations for uplink WIT with CoMP, denoted as $\{\tilde{\bold{q}}^{I*}_m\}$, are symmetric, which are obtained by separately considering the following two cases.
\begin{itemize}
\item If $D\le2H/\sqrt{3}$, the optimal hovering locations of the two UAVs are given as ${\bold{q}}^{I}_1(t)=\tilde{\bold{q}}^{I*}_1=(-d_{\text{min}}/2,0)$ and ${\bold{q}}^{I}_2(t)=\tilde{\bold{q}}^{I*}_2=(d_{\text{min}}/2,0)$;
\item If $D>2H/\sqrt{3}$, the optimal hovering locations of the two UAVs are
 given as ${\bold{q}}^{I}_1(t)=\bar{\bold{q}}^{I*}_1=(-\epsilon,0)$ and ${\bold{q}}^{I}_2(t)=\bar{\bold{q}}^{I*}_2=(\epsilon,0)$, $\forall t\in(\tilde{\tau}_E,T]$, where
 \begin{align}
 \epsilon\triangleq\max\left(\sqrt{-(D^2/4+H^2)+\sqrt{D^4/4+H^2D^2}},d_{\text{min}}/2\right)\le\max\left(d_{\min}/2,D/2\right).\nonumber \end{align}
\end{itemize}
\end{lemma}
\begin{IEEEproof}
Under given $\{Q_k^{\text{CoMP}}\}$, to obtain the UAVs' optimal hovering locations in problem (\ref{P62sum}), it is equivalent to solve following problem.
\begin{align}
\max_{|\tilde{x}^I|\ge d_{\min}/2}&~\frac{Q_k^{\text{CoMP}}\beta_0/\sigma^2} {(\tilde{x}^I+D/2)^2+H^2}+\frac{Q_k^{\text{CoMP}}\beta_0/\sigma^2} {(\tilde{x}^I-D/2)^2+H^2},\nonumber
\end{align}
which can be solved by checking the first-order derivative of the objective function. Therefore, Lemma \ref{lem4.1} is proved.
\end{IEEEproof}

Lemma \ref{lem4.1} shows that when the two IoT-devices are close, both UAVs should hover as close to the center of the two IoT-devices as possible with a minimum separation $d_{\min}$; otherwise, the two UAVs should hover at two symmetric locations between IoT-devices to enhance the common throughput. By substituting $\{\tilde{\bold{q}}^{I*}_m\}$ and $\{Q_k^{\text{CoMP}}\}$ into (\ref{P62}), we have the optimal common throughput of the two IoT-devices, denoted as $R_c^*(\tilde{\tau}_E)$.
\subsubsection{Complete Algorithm for Solving (P2.3)}
Finally, it remains to find the optimal $\tilde{\tau}_{E}^*$, which can be obtained via a 1D exhaustive search over $(0,T]$ to maximize the correspondingly achievable optimal data-rate throughput $R_c^*(\tilde{\tau}_E)$. Under $\tilde{\tau}_{E}^*$, the corresponding optimal data-rate throughput and optimal hovering locations become the obtained solution to problem (P2.3) or (P2.2). As a result, problem (P2.3) is solved.
\begin{remark}\label{Proposition:4.1}
The obtained solution to problem (P2.2) has a multi-location-hovering structure as follows. During downlink WPT period, the UAVs hover at symmetric locations between the two IoT-devices with $\|\tilde{\bold{q}}^{E,k*}_m\|\le\max(D/2,d_{\min}/2),\forall k,m\in\{1,2\}$; while during uplink WIT period, the UAVs hover at a different set of hovering locations given in Lemma \ref{lem4.1} that are above the center between the two IoT-devices with $\|\tilde{\bold{q}}^{I*}_m\|\le\max(D/2,d_{\min}/2),\forall m\in\{1,2\}$. This is in sharp contrast to the interference coordination case, in which during WIT, the UAVs generally need to fly far away from both IoT-devices for interference suppression.
\end{remark}

\vspace{-15pt}\subsection{Proposed Solution to Problem (P2.1) with Finite $T$}\label{sec8}
In this section, we propose an efficient solution to problem (P2.1) for the general case with finite $T$ based on the technique of alternation optimization. Similar to Section \ref{sec:V}, the time allocation $\{\rho_I[n],\rho_{E,k}[n]\}$, the IoT-devices' transmit power $\{Q_k[n]\}$, and the UAVs' trajectories $\{\bold{q}_m[n]\}$ are optimized sequentially in an alternating manner, with all other variables fixed.

\subsubsection{Time and Transmit Power Allocation}
For any given UAVs' trajectories $\{\bold{q}_m[n]\}$ and IoT-devices' transmit power $\{Q_k[n]\}$, the time allocation problem is expressed as
\begin{align}
\text{(P2.1-Time):}&\max_{\{\rho_I[n],\rho_{E,k}[n]\},R_c}~R_c\nonumber\\
&\mathrm{s.t.}~\text{(\ref{con69}), (\ref{P5c2}), (\ref{P5c3}), and (\ref{P2.1con}),}\nonumber
\end{align}
which is an LP. On the other hand, for any given UAVs' trajectories $\{\bold{q}_m[n]\}$ and time allocation $\{\rho_I[n],\rho_{E,k}[n]\}$, the IoT-devices' transmit power allocation problem becomes
\begin{align}
\text{(P2.1-Power):}&\max_{\{Q_k[n]\},R_c}~R_c\nonumber\\
&\mathrm{s.t.}~\text{(\ref{con69}) and (\ref{P2.1con}),}\nonumber
\end{align}
which is a convex optimization problem. As a result, both the time allocation problem (P2.1-Time) and the transmit power allocation optimization problem (P2.1-Power) can be optimally solved by standard convex optimization tools, such as CVX\cite{CVX}.

\subsubsection{UAVs' Trajectories Optimization}
Under any given IoT-devices' transmit power $\{Q_k[n]\}$ and time allocation $\{\rho_I[n],\rho_{E,k}[n]\}$, the UAVs' trajectories optimization problem is non-convex due to the non-convex constraints in (\ref{conconcon}), (\ref{con69}), and (\ref{P2.1con}). First, we introduce two variables $a_{k,m}[n]$ and $b_{k,m}[n]$, where $k\in\{1,2\},m\in\{1,2\}, n\in{\mathcal N}$. The UAVs' trajectories optimization problem can be reformulated as
\begin{align}
&\text{(P2.1-Tra):}\max_{\{a_{k,m}[n],b_{k,m}[n],\bold{q}_m[n]\},R_c}~R_c\nonumber\\
&\mathrm{s.t.}~\frac{1}{T}\sum_{{n}=1}^{{N}} \rho_I[{n}]\log_2\left(1+\frac{Q_k[{n}]\beta_0}{2\sigma^2}\sum_{m=1}^2 b_{k,m}[n]\right)\ge R_c,\forall k\in\{1,2\}\\
&\sum_{{n}=1}^{{N}}\left(\eta P\rho_{E,k}[{n}]\left(\sum_{m=1}^2a_{k,m}[{n}]\right)^2+\rho_{E,\bar k}[{n}]{E}'_{k}(\{\bold{q}_m[n]\})\right)\ge\sum_{{n}=1}^{{N}}Q_k[{n}]\rho_{I}[{n}],\forall k\in\{1,2\}\\
&~~~~~(a_{k,m}[{n}])^2\le \frac{\beta_0}{\lVert{\bold{q}}_m[{n}]-\bold{w}_k\rVert^2+H^2},\forall k,m\in\{1,2\}\\
&~~~~~\lVert{\bold{q}}_m[{n}]-\bold{w}_k\rVert^2+H^2\le\frac{1}{b_{k,m}[ n]},\forall k,m\in\{1,2\},n\in{\mathcal N}\\
&~~~~~\text{(\ref{IFC}), (\ref{speedcon}), and (\ref{conconcon}).}\nonumber
\end{align}
Problem (P2.1-Tra) is still non-convex. To tackle these non-convex constraints, we use SCA to update the UAVs' trajectories $\{{\bold{q}}_m[{n}]\}$, $\{a_{k,m}[n]\}$, and $\{b_{k,m}[n]\}$ in an iterative manner. Let $\{\bold{q}_m^{(l)}[n]\},\{a_{k,m}^{(l)}[{n}]\}$, and $\{b_{k,m}^{(l)}[ n]\}$ denote the local points at the $l$-th iteration. In addition to using the lower bound approximations in (\ref{lbr}), (\ref{lbe}), and (\ref{lbq}), we have the lower bounds of $(\sum_{m=1}^2a_{k,m}[{n}])^2$ and $\frac{1}{b_{k,m}[ n]}$ as follows, respectively, by obtaining their first-order Taylor expansions.

\vspace{-1em}\begin{footnotesize}
\begin{align}
&(\sum_{m=1}^2a_{k,m}[{n}])^2\ge(a_{k,1}^{(l)}[ n]+a_{k,2}^{(l)}[n])^2+2(a_{k,1}^{(l)}[n]+a_{k,2}^{(l)}[ n])(a_{k,1}[ n]+a_{k,2}[ n]-a_{k,1}^{(l)}[ n]-a_{k,2}^{(l)}[ n]),\\
&\frac{1}{b_{k,m}[ n]}\ge\frac{1}{b_{k,m}^{(l)}[ n]}-\frac{1}{(b_{k,m}^{(l)}[n])^2}(b_{k,m}[ n]-b_{k,m}^{(l)}[ n]),\forall k,m\in\{1,2\},n\in\mathcal N.
\end{align}\end{footnotesize}

\vspace{-1em}As a result, problem (P2.1-Tra) can be approximated into a convex optimization problem, which can be efficiently solved by CVX. Therefore, an efficient solution to the trajectory optimization problem in (P2.1-Tra) is obtained.

In order to efficiently implement the alternating optimization based algorithm, we choose the SHF trajectory as the initial trajectory for iteration. In particular, in the considered SHF trajectory, the two UAVs sequentially visit their respective initial locations, hovering locations for IoT-device 1's downlink WPT, uplink WIT, and IoT-device 2's downlink WPT, respectively, and final locations, at the maximum flying speed. Furthermore, let $\tilde{T}_{\text{fly}}$ denote the minimum flying duration for UAVs to visit these locations. If the total duration $T<\tilde{T}_{\text{fly}}$, then we simply choose the direct flight trajectory design as that in Section \ref{sec:V}, under which the time and power allocation can be obtained by solving problems (P2.1-Time) and (P2.1-Power), respectively.

In summary, problem (P2.1) is solved via solving three subproblems in an alternating manner. Such an alternating optimization ensures the objective value of problem (P2.1) to be monotonically non-decreasing after each iteration. Thus, the proposed design for problem (P2.1) is guaranteed to converge.

\vspace{-20pt}\section{Numerical Results}\label{sec9}
In the simulation, the UAVs fly at a fixed altitude of $H=5$~m. The receiver noise power at the UAVs is set as $\sigma^2=-100$~dBm. The channel power gain at the reference distance $d_0=1$~m is set as $\beta_0=-30$~dB. The UAVs' transmit power is $P=40$~dBm. The energy harvesting efficiency is set as $\eta=60\%$. The maximum flying speed of the UAVs is $\tilde{V}_{\text{max}}=5$~m/s. The minimum inter-UAV distance is $d_{\text{min}}=1$~m.
%
\begin{figure}
\centering
\subfigure[The interference coordination scenario with $D=5$~m]
{
    \begin{minipage}{7.5cm}
    \centering
    \includegraphics[width=7cm]{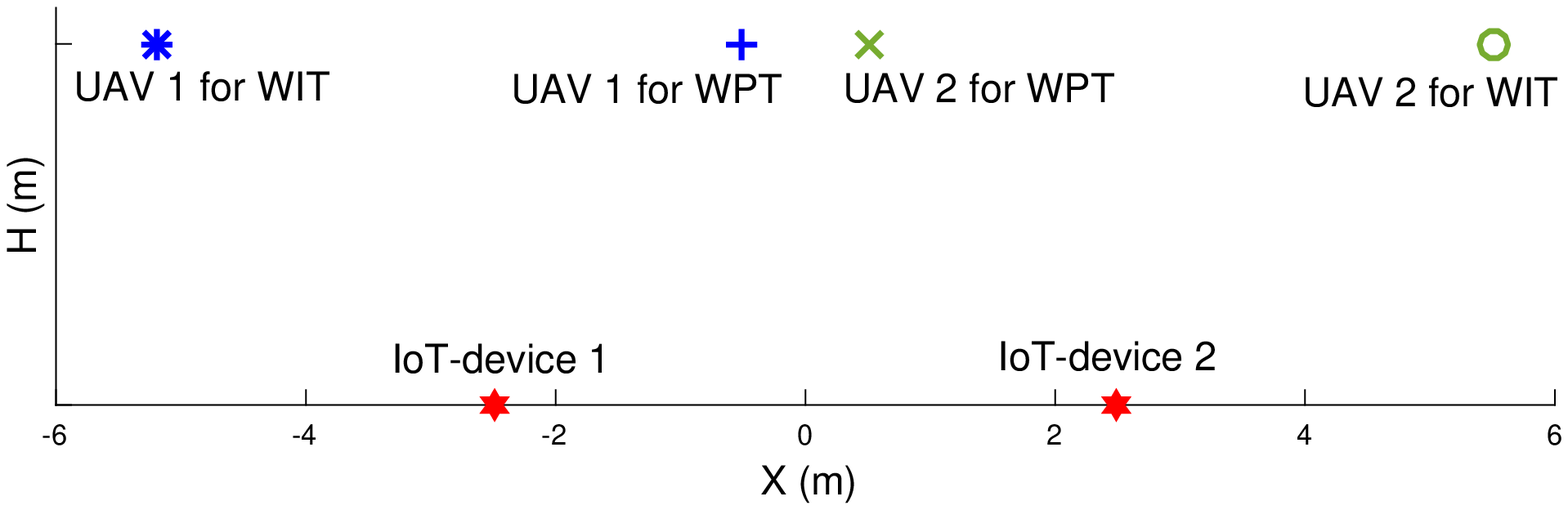}
    \end{minipage}
 }
 \subfigure[The interference coordination scenario with $D=15$~m]
{
    \begin{minipage}{7.5cm}
    \centering
    \includegraphics[width=7cm]{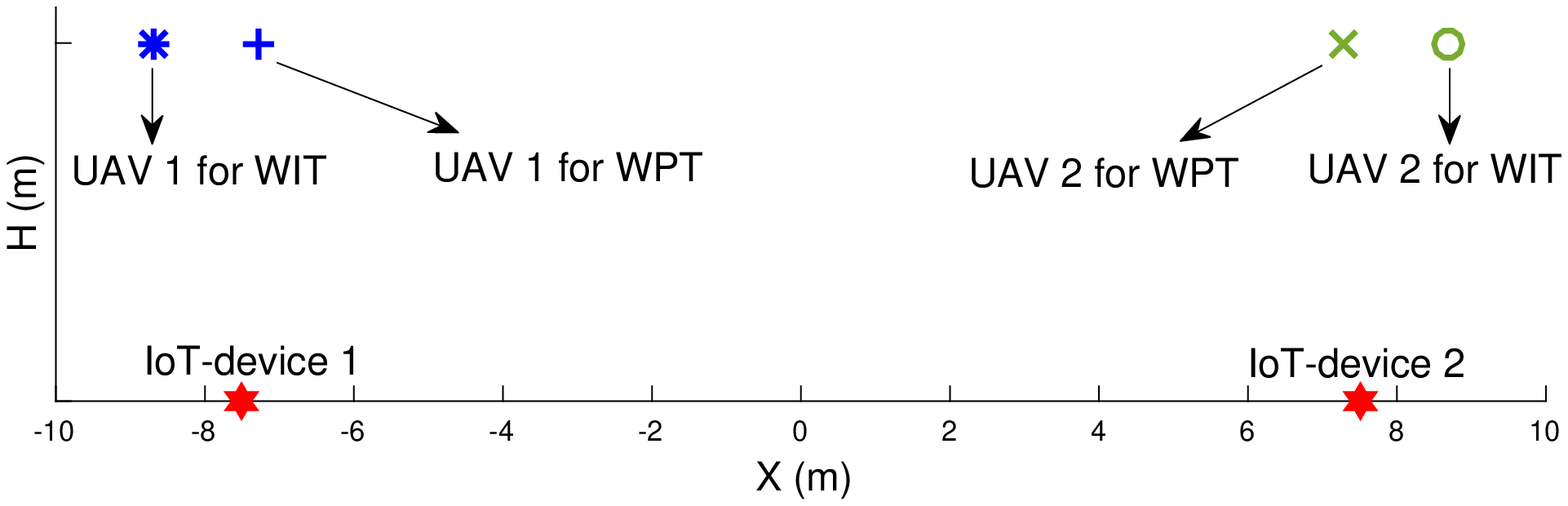}
    \end{minipage}
 }
  \subfigure[The CoMP scenario with $D=5$~m]
{
    \begin{minipage}{7.5cm}
    \centering
    \includegraphics[width=7cm]{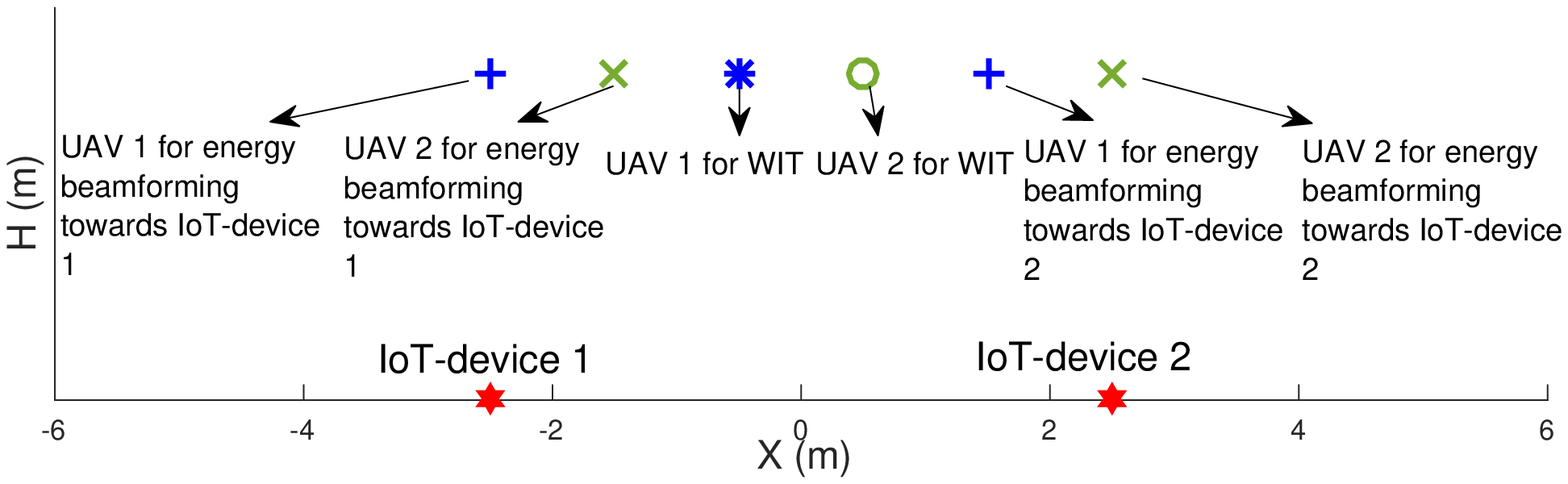}
    \end{minipage}
 }
  \subfigure[The CoMP scenario with $D=15$~m]
{
    \begin{minipage}{7.5cm}
    \centering
    \includegraphics[width=7cm]{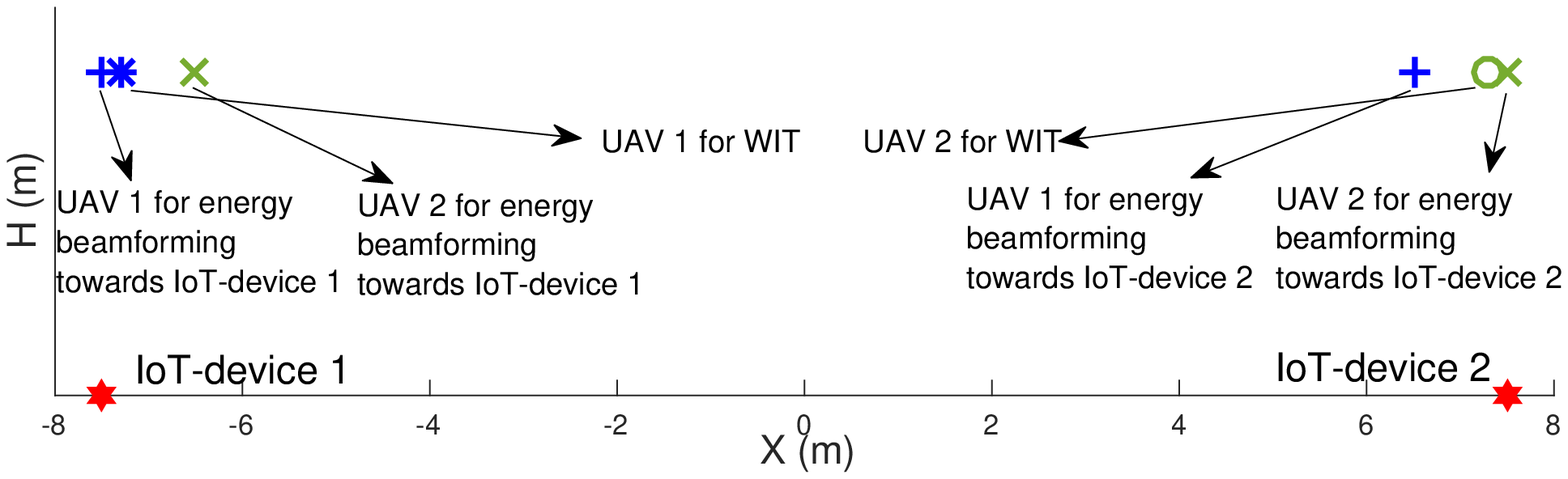}
    \end{minipage}
 }

\caption{The UAVs' optimal hovering locations under the interference coordination and CoMP scenarios, with different values of the IoT-devices' distance $D$. Here, each UAV has more than one hovering locations: In (a) and (b) each UAV has 2 hovering locations each for WIT and WPT, respectively; in (c) and (d) each UAV has 3 hovering locations, two for energy beamforming towards IoT-devices 1 and 2, and one for WIT, respectively. These UAVs stay at these locations at different time.} \label{fig:2}
\end{figure}

First, we consider the special case when the UAV mission duration $T\to\infty$, Fig. \ref{fig:2} shows the UAVs' optimal hovering locations for WIT and WPT, in which the distance between the two IoT-devices is set as $D=5$~m and $D=15$~m. It is observed that the optimal hovering locations of both UAVs are symmetric. In the interference coordination scenario, each UAV has 2 hovering locations, each for WIT and WPT, respectively, where the hovering locations for WIT are far away from the IoT-devices while those for WPT are located between the two IoT-devices, which is in accordance with Remark \ref{Remark:3.2}. Moreover, in the CoMP scenario, each UAV has 3 hovering locations, two for energy beamforming towards IoT-devices 1 and 2, and one for WIT, respectively. This is consistent with the results in Section \ref{sec6}. It is observed that the two UAVs successively hover above the IoT-devices for downlink WPT with collaborative energy beamforming, and hover between the two IoT-devices for uplink WIT with CoMP, which is in accordance with Remark \ref{Proposition:4.1}.
\begin{figure}
\begin{minipage}[t]{0.5\textwidth}
\centering
    \includegraphics[width=6cm]{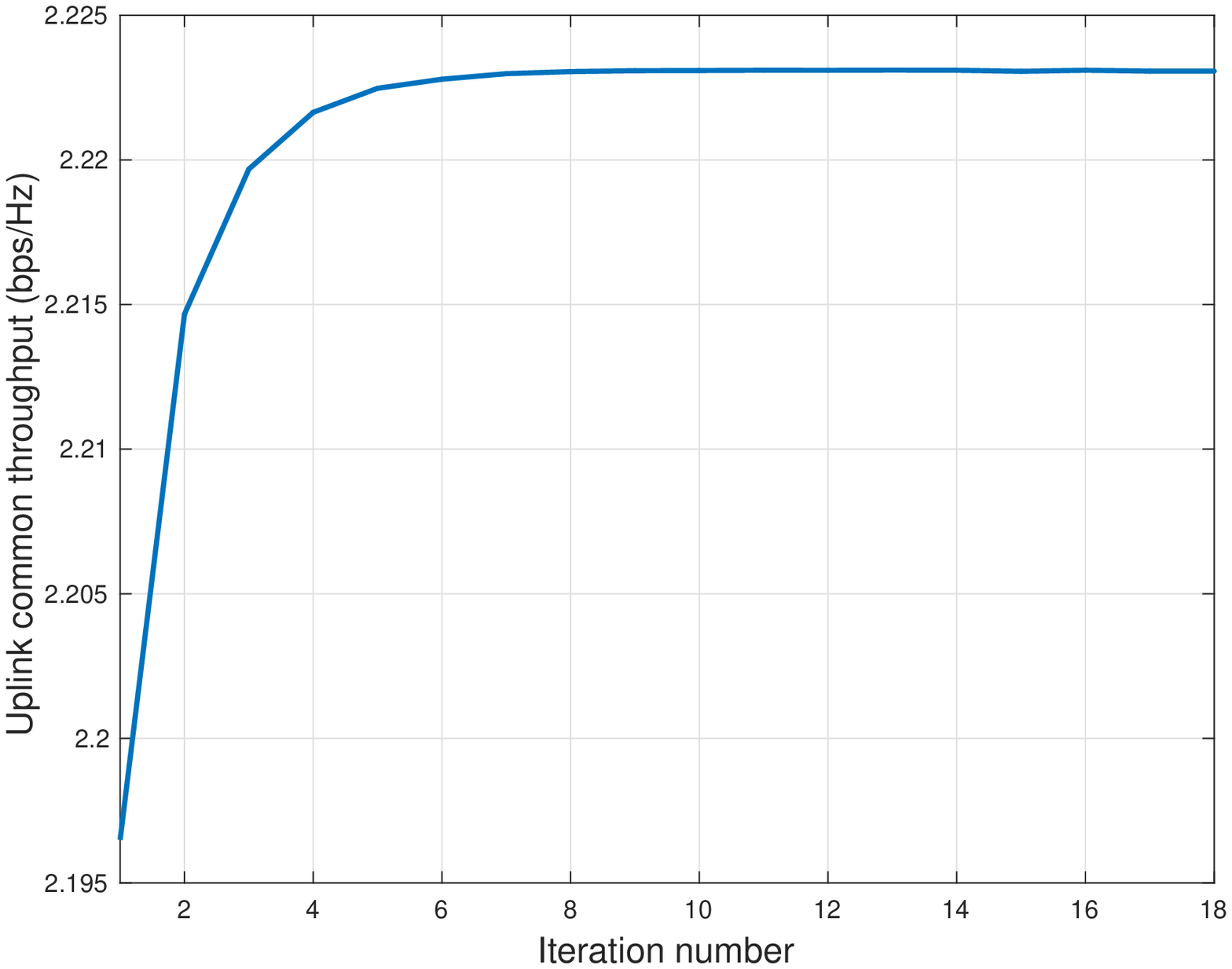}
\caption{Convergence behavior of the alternative optimization based approach for solving (P1).} \label{fig:initer}
\end{minipage}~
\begin{minipage}[t]{0.5\textwidth}
\centering
    \includegraphics[width=6cm]{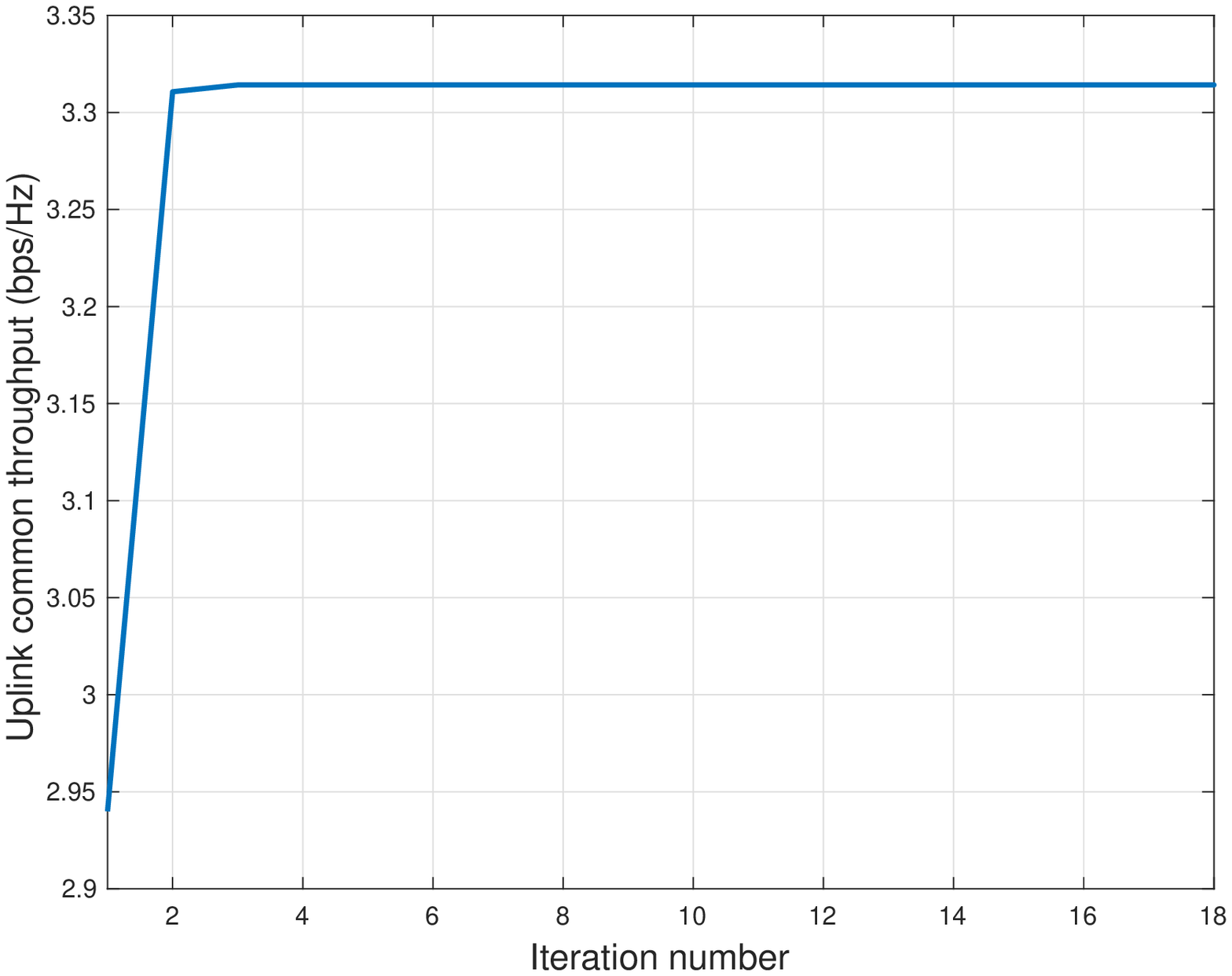}
\caption{Convergence behavior of the alternative optimization based approach for solving (P2.1).} \label{fig:compiter}
\end{minipage}
\end{figure}
\begin{figure}
\centering
\subfigure[Two UAVs' trajectories in interference coordination scenario]
{
    \begin{minipage}{5cm}
    \centering
    \includegraphics[width=5.5cm]{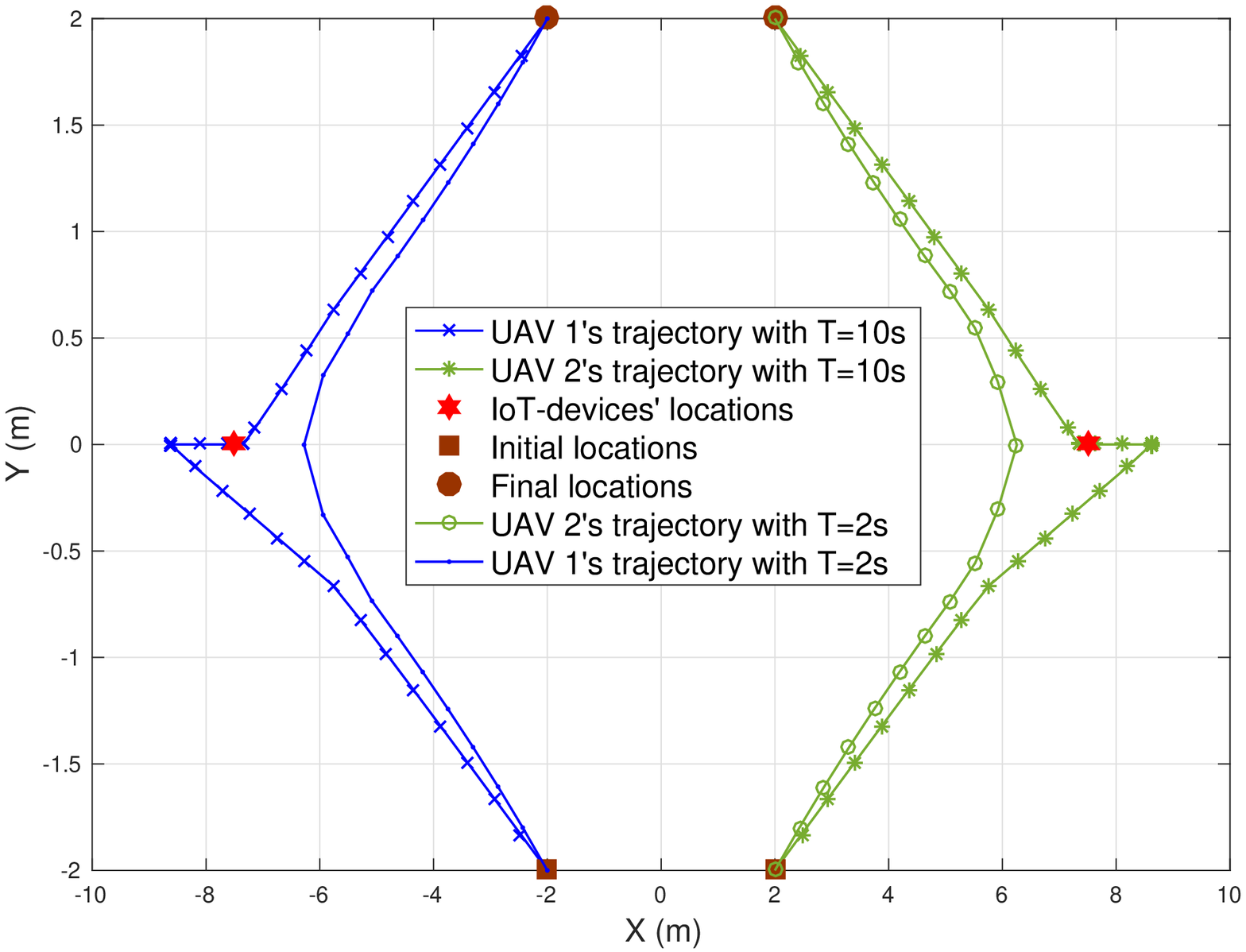}
    \end{minipage}
 }
 \subfigure[UAV 1's trajectory in CoMP scenario]
{
    \begin{minipage}{5cm}
    \centering
    \includegraphics[width=5.5cm]{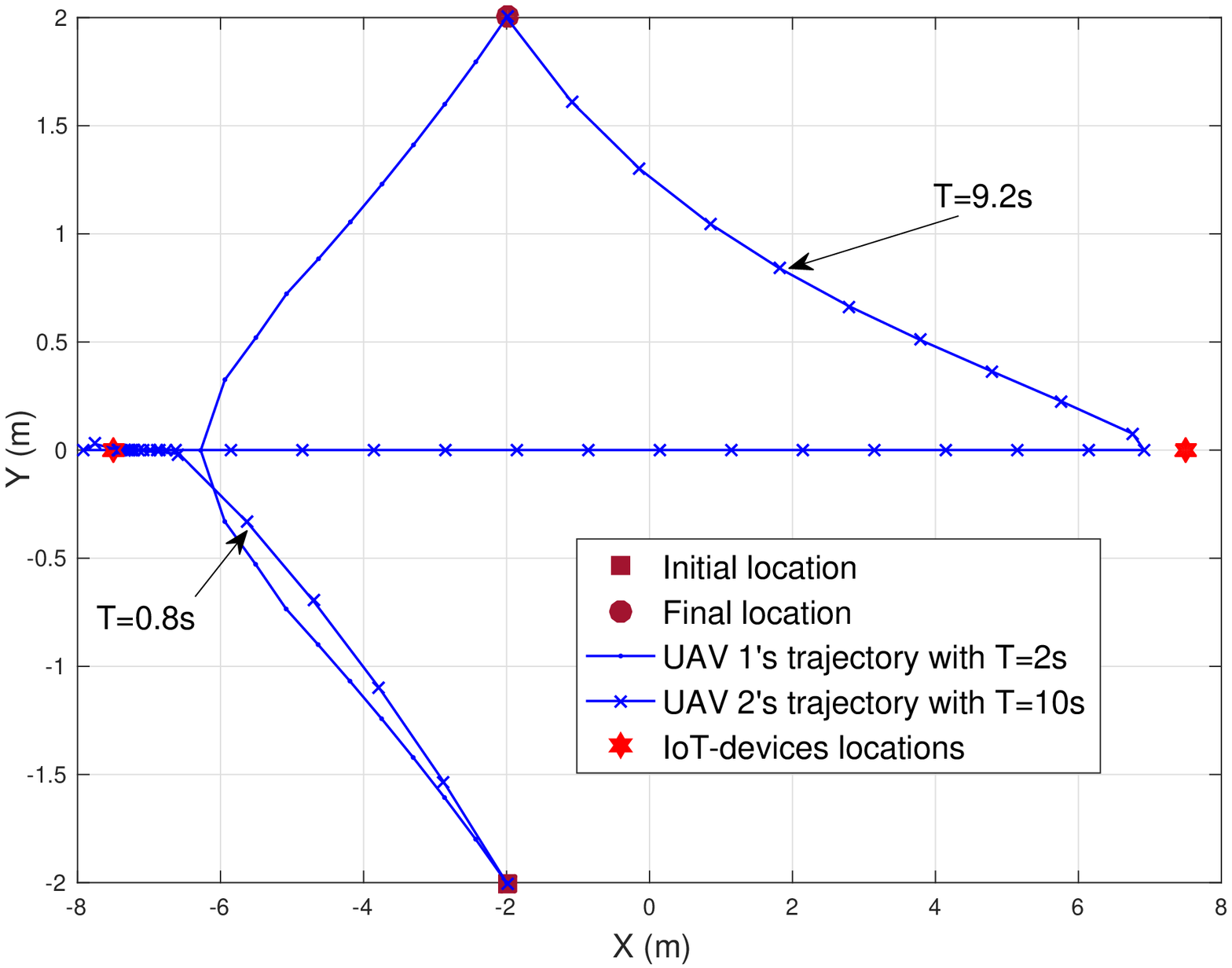}
    \end{minipage}
 }
  \subfigure[UAV 2's trajectory in CoMP scenario]
{
    \begin{minipage}{5cm}
    \centering
    \includegraphics[width=5.5cm]{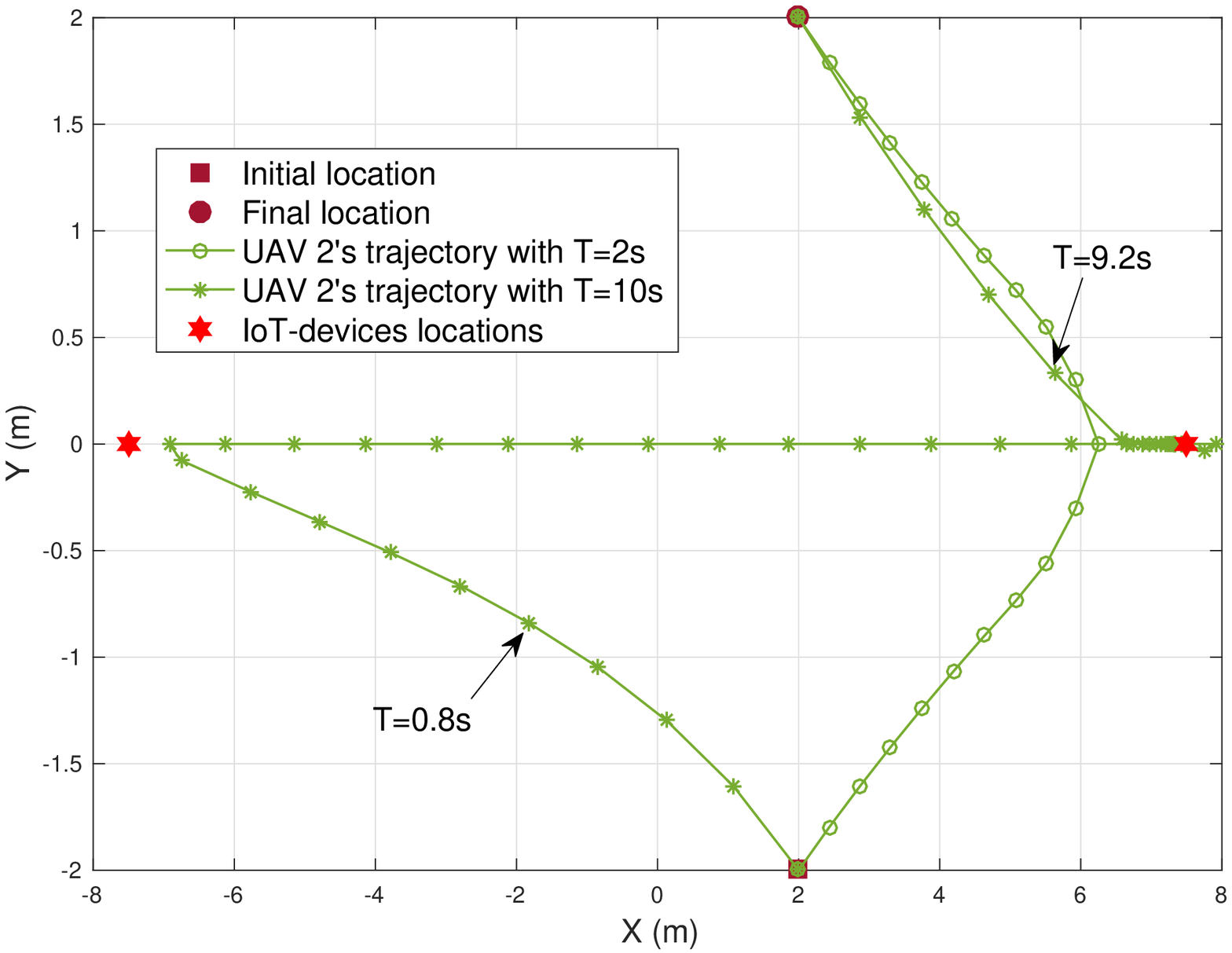}
    \end{minipage}
 }

\caption{Obtained UAVs' trajectories in different scenarios with $D=15$~m.} \label{fig:5}
\end{figure}

Next, we show the performance of our proposed joint trajectory and wireless resource allocation design for both scenarios of interference coordination and CoMP. Figs. \ref{fig:initer} and \ref{fig:compiter} show the convergence behavior of the alternating-optimization-based algorithms for solving problems (P1) and (P2.1), respectively, where $D=15$~m and $T = 4$~s. It is observed that in both scenarios, the uplink common throughput are monotonically increasing after each iteration, and the algorithms converge very fast. Next, we consider the following benchmark scheme for performance comparison, which is intuitively designed and has been used in other setups as benchmarks (see, e.g., \cite{ShuoWen}).
\begin{itemize}
	\item {\bf{Direct flight}}: The two UAVs directly fly from the initial locations to final locations with straight path and constant speed. In this case, the uplink common throughput maximization problem is reduced to transmission allocation problem with fixed trajectory, which can be solved by the corresponding algorithms presented in Sections \ref{sec:P2} and \ref{sec6} for the interference coordination and CoMP scenarios, respectively.
\end{itemize}

Fig. \ref{fig:5} shows the obtained UAVs' trajectories under the proposed designs, where $T=10$~s and $T=2$~s, respectively. It is observed that when $T=10$~s, the two UAVs try to reach the optimized hovering locations (that derived for the case with infinite $T$ as shown in Fig. \ref{fig:2}), in order to achieve efficient WIT and WPT. For the CoMP scenario in this case, both UAVs are observed to visit similar locations at different time instants, thus avoiding the UAVs' collision. It is further observed that when $T = 2$~s, the two UAVs' trajectories become similar for both scenarios, which deviate from the optimal hovering locations due to the limited mission duration available.

\begin{figure}
\begin{minipage}[t]{0.5\textwidth}
\centering
    \includegraphics[width=8cm]{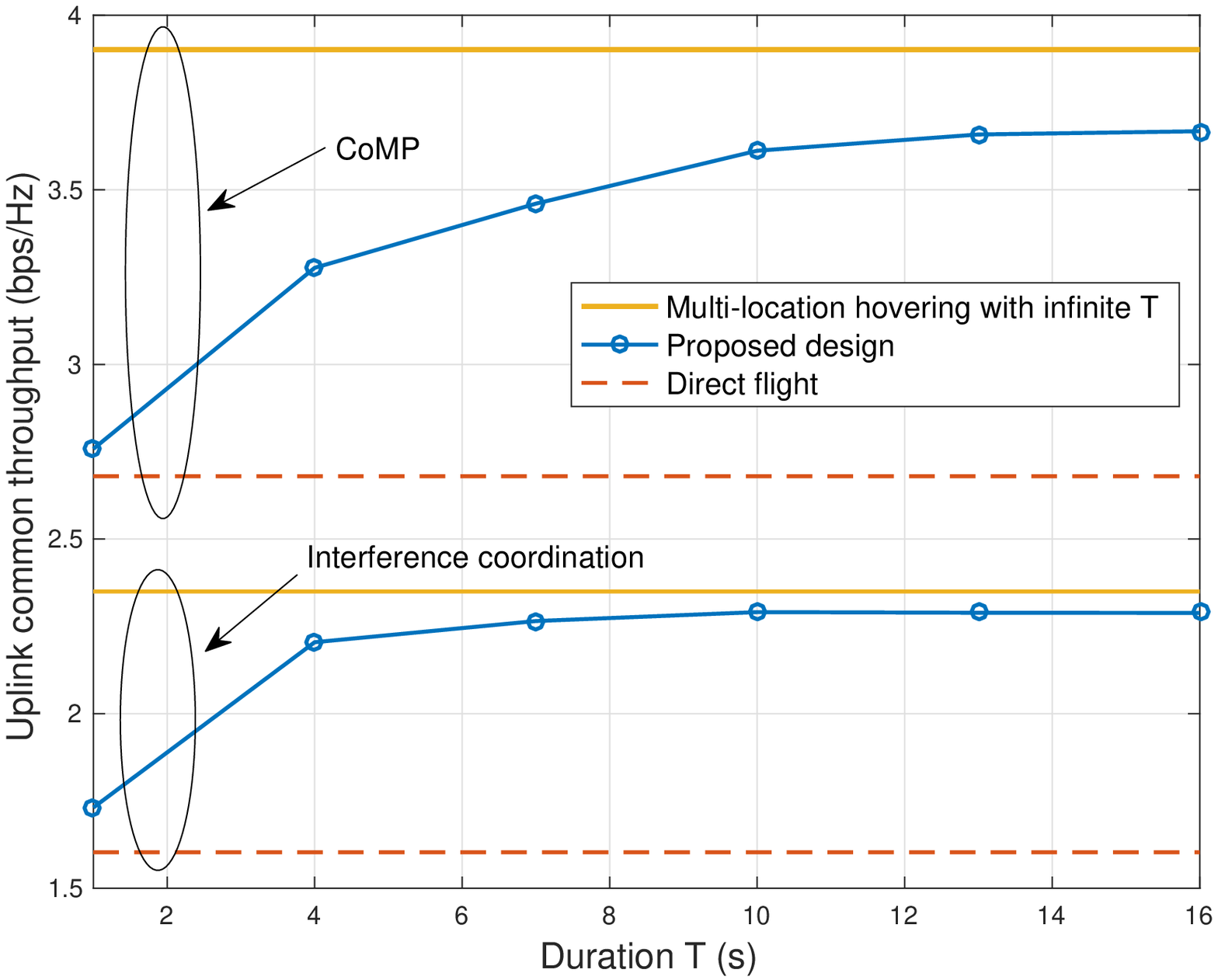}
\caption{The uplink common throughput versus the duration $T$ under different scenarios with $D=15$~m.} \label{fig:7}
\end{minipage}~
\begin{minipage}[t]{0.5\textwidth}
\centering
    \includegraphics[width=8cm]{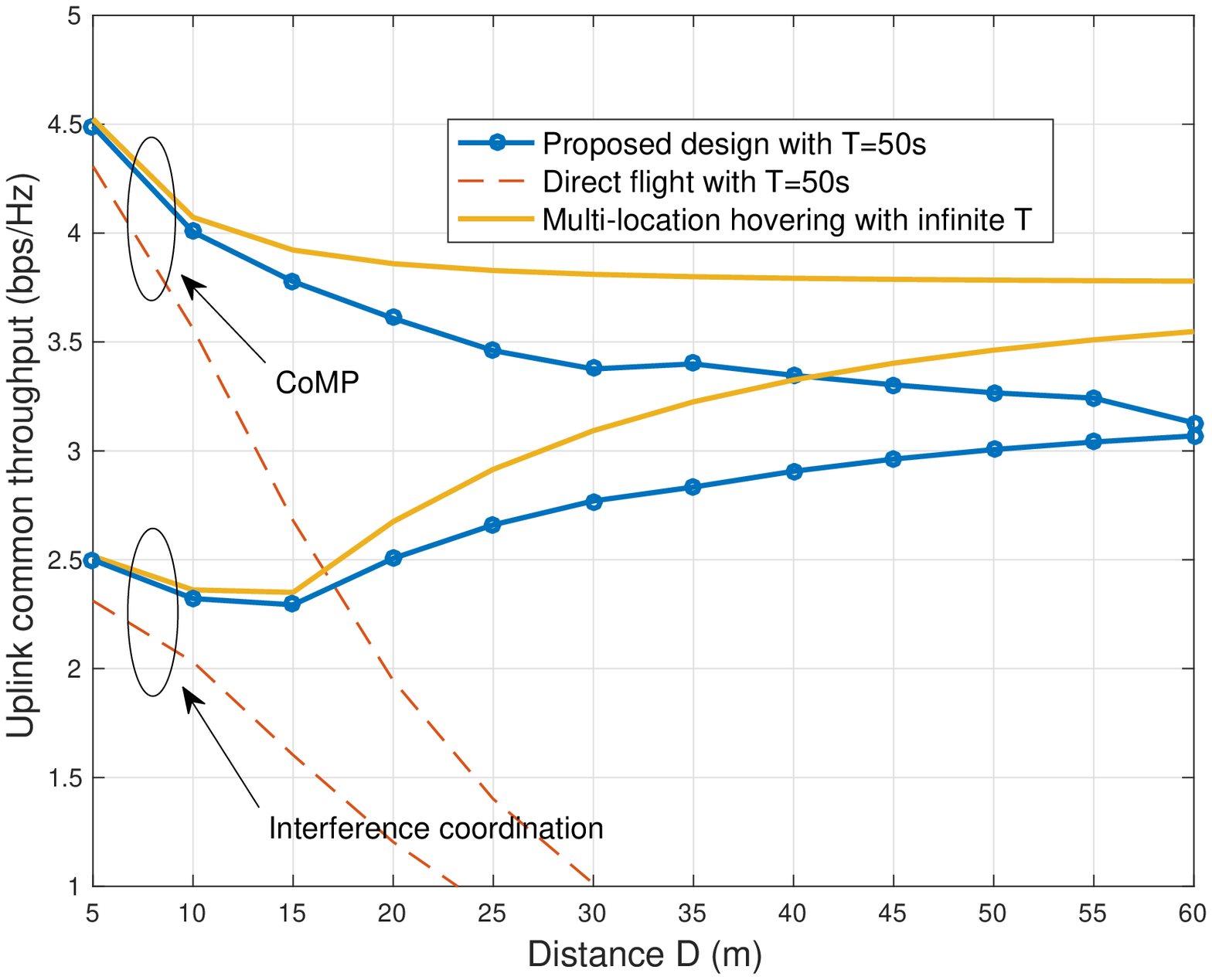}
\caption{The uplink common throughput versus the IoT-devices' distance $D$ under different scenarios with $T=50$~s.} \label{fig:12}
\end{minipage}
\end{figure}

Fig. \ref{fig:7} shows the uplink common throughput of the two IoT-devices versus the UAV mission duration $T$ with $D=15$~m. It is observed that for both scenarios with interference coordination and CoMP, the proposed joint trajectory and wireless resource allocation design achieves much higher common throughput than the direct flight benchmark, and the performance gain becomes more substantial as $T$ increases. Furthermore, when $T$ becomes sufficiently large, the proposed design is able to approach the performance upper bound by the multi-location-hovering solution to (P1.1) or (P2.2), respectively. In addition, the employment of CoMP transmission/reception at the UAVs can significantly enhance the uplink common throughput performance.

Fig. \ref{fig:12} shows the uplink common throughput versus the distance $D$ between the two IoT-devices, with IoT-devices' locations $(-D/2,0)$ and $(D/2,0)$, and UAVs' initial and final locations being (-2~m, -2~m) and (-2 m, 2 m) for UAV 1, as well as (2~m, -2~m) and (2~m, 2~m) for UAV 2, respectively. It is observed that the employment of CoMP always outperforms interference coordination, and our proposed design always outperforms the benchmark direct flight trajectory design. Specifically, in the interference coordination scenario, as $D$ increases, the uplink common throughput for the proposed design is observed to first decrease and then increases. This is due to the fact that in this case the system first operates in the TDMA transmission mode then in the simultaneous transmission mode, as shown in Fig. \ref{fig:10}. Moreover, as $D$ increases to a sufficiently large value, the performance gap between CoMP and interference coordination becomes smaller, as the cross channels between UAVs and IoT-devices become weaker, thus compromising the cooperation gain brought by CoMP. In this case, employing the MMSE-beamforming for WIT under the optimized UAV trajectory (under ZF-beamforming) is expected to further enhance the system performance, which, however, is left for future work.

%

\vspace{-20pt}\section{Concluding Remarks}\label{sec:con}
In this paper, we investigated the uplink common throughput maximization in a UAV-enabled two-user interference channel with WPCN by jointly optimizing both UAVs' cooperative trajectories design and wireless resource allocations, subject to the UAVs' maximum flying speed and collision avoidance constraints, as well as the IoT-devices' individual energy neutrality constraints. By considering two scenarios, namely interference coordination and CoMP transmission/reception, we presented  optimal multi-location-hovering  solutions in the special case when the UAV mission duration is sufficiently long, which show that  the UAVs should hover at different sets of locations for downlink WPT and uplink WIT, respectively. In the general case with finite UAV mission duration, we obtained suboptimal but high-quality solutions by using the techniques of alternating optimization and SCA. Numerical results showed that our proposed designs significantly enhance the WPCN performance, and the CoMP design leads to higher  throughput than the interference coordination at the cost of higher implementation complexity. Due to the space limitation, there are various interesting issues remaining unaddressed in this paper, which are briefly discussed in the following to motivate future work.
\begin{itemize}
	\item This paper considered that the two IoT-devices are deployed at fixed locations that are {\it a-priori} known by UAVs, in which case the association between UAVs and IoT-devices is predetermined. In practice, the devices can also be mobile with locations varying over time. Accordingly, the UAV-device association may need to be adaptively optimized jointly with the UAV trajectory design and wireless resource allocation for performance enhancement. Furthermore, as the devices' moving locations may not be fully predictable over time, how to design the online UAV trajectory design and resource management in real time via e.g. dynamic programming and (deep) reinforcement learning is an interesting problem worth investigation in future work.
	\item Furthermore, this paper assumed simplified LoS-dominated wireless channel between UAVs and IoT-devices for the purpose of initial investigation. Under more complicated environment in practice, there may exist obstacles between UAVs and IoT-devices, thus making the channel propagation properties more complex. How to optimize the UAV trajectory design and wireless resource allocation under such non-LoS and unsteady channels over both time and space is another interesting problem that is difficult to solve. Radio map\cite{xiaopengmo} and reinforcement learning may be viable tools to provide efficient solutions.
	\item This paper only considered two UAVs serving two IoT-devices for the purpose of initial investigation, while there may exist a large number of UAVs and IoT-devices in practical large-scale WPCNs. In this case, how to design the UAV trajectory and wireless resource allocation becomes more difficult, due to the involvement of more design variables and more complicated interference environments. One feasible solution may be to first cluster IoT-devices into different groups and design the collaborative energy beamforming and interference mitigation/exploitation within each group  respectively.
\end{itemize}

\vspace{-20pt}\appendix
\vspace{-15pt}\subsection{Proof of Lemma \ref{lem3.1}}\label{AppA}
For notational convenience, denote the objective function of problem (\ref{P31}) as $\phi(x^E)$. The first-order derivative of $\phi(x^E)$ is obtained as
\begin{align}
\phi'(x^E)=-4\eta\tau_E\beta_0 Px^E\left(\frac{(x^E)^4+2(D^2/4+H^2)(x^E)^2-3D^4/16+H^4-H^2D^2/2}{((x^E)^2+D^2/4+H^2-Dx^E)^2((x^E)^2+D^2/4+H^2+Dx^E)^2}\right).\label{931}
\end{align}
Similarly as in \cite{JieXuWPT}, it can be easily shown that when $D\le 2H/\sqrt{3}$, we have $\phi'(x^E)\le0$ for any $x^E\ge0$. As a result, $\phi(x^E)$ is monotonically non-increasing over $x^E\in[0,\infty )$. In the other case with $D>2H/\sqrt{3}$, it can be shown that $\phi'(x^E)\ge0$ for $x^E\in[0,\epsilon )$, and $\phi'(x^E)\le0$ for $x\in[\epsilon,\infty )$. As a result, $\phi(x^E)$ is monotonically non-decreasing and non-increasing over $x\in[0,\epsilon )$ and $x\in[\epsilon,\infty )$, respectively. By considering the collision avoidance constraint $x^E\ge d_{\text{min}}/2$, we have that if $D\le 2H/\sqrt{3}$, then the optimal solution to problem (\ref{P31}) is $x^{E*} = d_{\min}/2$; while if $D> 2H/\sqrt{3}$, then $x^{E*} =\epsilon$.

By substituting $x^{E*}$ into problem (\ref{P1.1.3}), Lemma \ref{lem3.1} is finally proved.

\begin{small}

\end{small}

\end{document}